\documentclass[conference]{IEEEtran}

\usepackage[hidelinks]{hyperref}
\usepackage{xcolor,colortbl}
\usepackage{hhline}
\usepackage{amsmath}
\usepackage{multirow}
\usepackage{amsfonts}
\usepackage{caption}
\usepackage{tikz}
\usepackage{pgfplots}
\usepackage{booktabs}
\usepackage{pifont}
\usepackage{float}
\usepackage{tcolorbox}
\usepackage{tabularx}
\usepackage[flushleft]{threeparttable}
\usepackage[T1]{fontenc}
\usepackage[titlenumbered,ruled,noline,linesnumbered]{algorithm2e}
\usepackage{gensymb}
\usepackage{flushend}
\usepackage{makecell}
\usepackage{subfigure}

\usetikzlibrary{calc}
\usetikzlibrary{patterns}
\usetikzlibrary{shapes.multipart, arrows, positioning,shapes.geometric}
\usetikzlibrary{decorations.pathreplacing}
\usetikzlibrary{calc,shapes,decorations,decorations.shapes,positioning,shadows,arrows,decorations.markings,backgrounds,fit}

\makeatletter
\newcommand*{\rom}[1]{\expandafter\@slowromancap\romannumeral #1@}
\makeatother
\renewcommand{\vec}[1]{\mathbf{#1}}

\DeclareMathOperator*{\argmax}{arg\,max}

\begin{document}
  
\title{Defending Model Inversion and Membership Inference Attacks via Prediction Purification}

\author{\IEEEauthorblockN{Ziqi Yang\IEEEauthorrefmark{1}\IEEEauthorrefmark{4},
Bin Shao\IEEEauthorrefmark{1},
Bohan Xuan\IEEEauthorrefmark{1}, 
Ee-Chien Chang\IEEEauthorrefmark{4} and
Fan Zhang\IEEEauthorrefmark{1}\IEEEauthorrefmark{2}\IEEEauthorrefmark{3}}
\IEEEauthorblockA{\IEEEauthorrefmark{1}Zhejiang University\\Email: \{yangziqi, shaobin\_zju, xuanbohan, fanzhang\}@zju.edu.cn}
\IEEEauthorblockA{\IEEEauthorrefmark{2}Alibaba-Zhejiang University Joint Institute of Frontier Technologies}
\IEEEauthorblockA{\IEEEauthorrefmark{3}Zhejiang Lab}
\IEEEauthorblockA{\IEEEauthorrefmark{4}National University of Singapore\\Email: changec@comp.nus.edu.sg}}

\maketitle

\begin{abstract}

Neural networks are susceptible to data inference attacks such as the model inversion attack and the membership inference attack, where the attacker could infer the reconstruction and the membership of a data sample from the confidence scores predicted by the target classifier.
In this paper, we propose a unified approach, namely purification framework, to defend data inference attacks.
It purifies the confidence score vectors predicted by the target classifier by reducing their dispersion.
The purifier can be further specialized in defending a particular attack via adversarial learning.
We evaluate our approach on benchmark datasets and classifiers.
We show that when the purifier is dedicated to one attack, it naturally defends the other one, which empirically demonstrates the connection between the two attacks.
The purifier can effectively defend both attacks. For example, it can reduce the membership inference accuracy by up to 15\% and increase the model inversion error by a factor of up to 4.
Besides, it incurs less than 0.4\% classification accuracy drop and less than 5.5\% distortion to the confidence scores.

\end{abstract}

\section{Introduction}

Machine learning has been widely adopted in a variety of applications, transforming many aspects of our daily life such as handling users' sensitive data. Machine learning itself has also been provided as a service, e.g., machine-learning-as-a-service, by many platforms. Users access these models through prediction APIs which return a prediction score vector.  
Such vector is a probability distribution over the possible classes and each score indicates the confidence in predicting the corresponding class.
The class with the largest confidence is predicted as the label of the input data. In this paper, we are interested in black-box data inference attacks, notably \textit{membership inference} and \textit{model inversion} that exploit such prediction scores to threaten the security of machine learning.

A series of studies has indicated that the prediction scores of black-box machine learning models could be exploited to perform data inference attacks to get useful information about the data on which the machine learning model operates~\cite{shokri_membership_2017, nasr_comprehensive_2019, nasr_machine_2018, jia_memguard_2019, salem_ml-leaks_2018, yang_neural_2019, fredrikson_model_2015, hitaj_deep_2017}.
Membership inference attack~\cite{shokri_membership_2017, nasr_comprehensive_2019, nasr_machine_2018, jia_memguard_2019, salem_ml-leaks_2018} and model inversion attack~\cite{yang_neural_2019, fredrikson_model_2015, hitaj_deep_2017} are two of the most important and exemplary ones.
In a membership inference attack, the adversary is asked to determine whether a given data sample is in the target model's training data or not according to the prediction results of the target model. For example, the most common way is to train a binary classifier which takes the prediction scores as input and predicts whether the data sample is a member or non-member of the target model. 
In a model inversion attack, the adversary aims at inferring information about the data sample from the prediction results such as the reconstruction of the sample~\cite{yang_neural_2019, fredrikson_model_2015, hitaj_deep_2017} or the sensitive attributes~\cite{fredrikson_privacy_2014, wu_methodology_2016, hidano_model_2017}. Yang et al.~\cite{yang_neural_2019} proposed an effective black-box model inversion attack. The attacker leverages auxiliary knowledge to construct an inversion model which can reconstruct the original input sample from the prediction scores with high accuracy.

Although the membership inference attack and the model inversion attack aim at different goals, it is interesting to raise the following research questions: Are the two types of data inference attacks connected? Can they be mitigated in a unified approach? 
Unfortunately, the two attacks are primarily studied independently from each other in the literature. Furthermore, most previous defense approaches focused on the membership inference attack, while little has been studied about the defense of model inversion attack.

The major cause of membership inference attack is that the prediction scores of the target model are distinguishable for members and non-members of the training data~\cite{shokri_membership_2017, jia_memguard_2019}. Such distinguishability leaks membership information to the attacker. 
A number of approaches have been proposed to mitigate the membership inference attack in the literature.
They mainly aim at reducing overfitting by various regularization techniques, such as $L_2$ regularizer~\cite{shokri_membership_2017}, dropout~\cite{salem_ml-leaks_2018}, model-stacking~\cite{salem_ml-leaks_2018} and min-max regularization~\cite{nasr_machine_2018}.  
However, they do not impose a direct reduction of the distinguishability. 
Although overfitting is believed to be one of the major reasons causing the distinguishability, it is not the only reason~\cite{shokri_membership_2017}.
Jia et al.~\cite{jia_memguard_2019} proposed MemGuard which, instead of reducing overfitting, transforms the prediction scores to an adversarial example to evade the attacker's membership classification.
The defense performance relies on the transferability~\cite{papernot2016transferability} of adversarial examples, which does not essentially reduce the distinguishability as well. 
Another set of defenses use differential privacy mechanisms~\cite{abadi_deep_2016, iyengar_towards_2019, shokri_privacy-preserving_2015}. 
These approaches can provide a theoretical guarantee of privacy but impose a significant classification accuracy loss~\cite{nasr_machine_2018}.

\begin{figure}[t]
	\centering
	\begin{minipage}[b]{1\linewidth}
		\centering
		\subfigure[]{
			\includegraphics[width=0.46\linewidth]{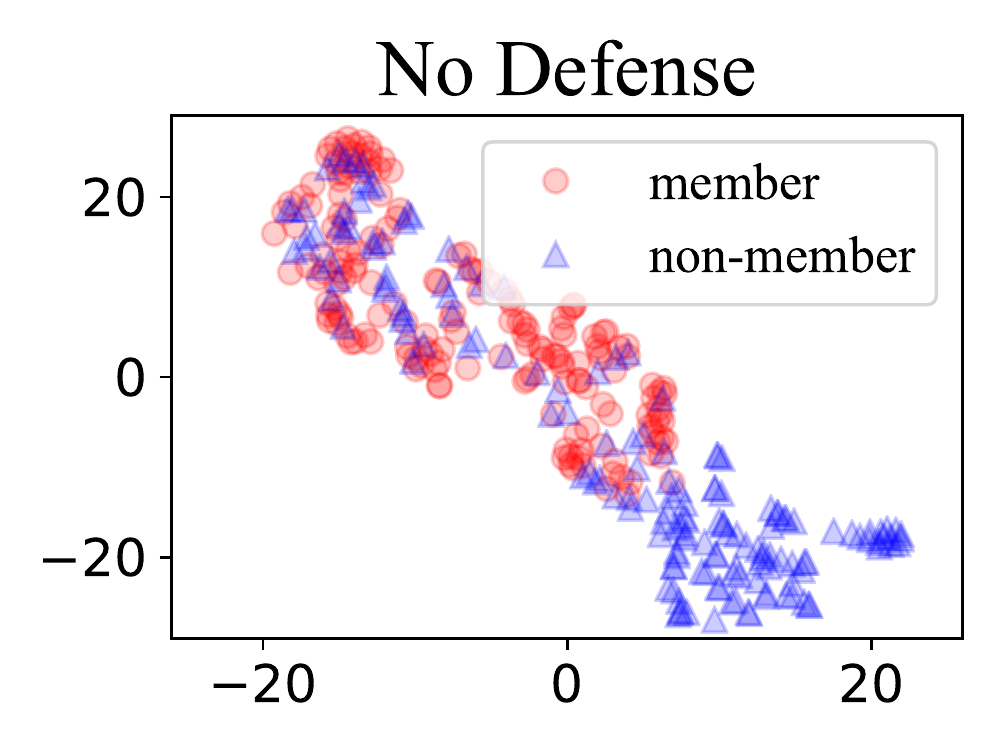}
		}
		\subfigure[]{
			\includegraphics[width=0.46\linewidth]{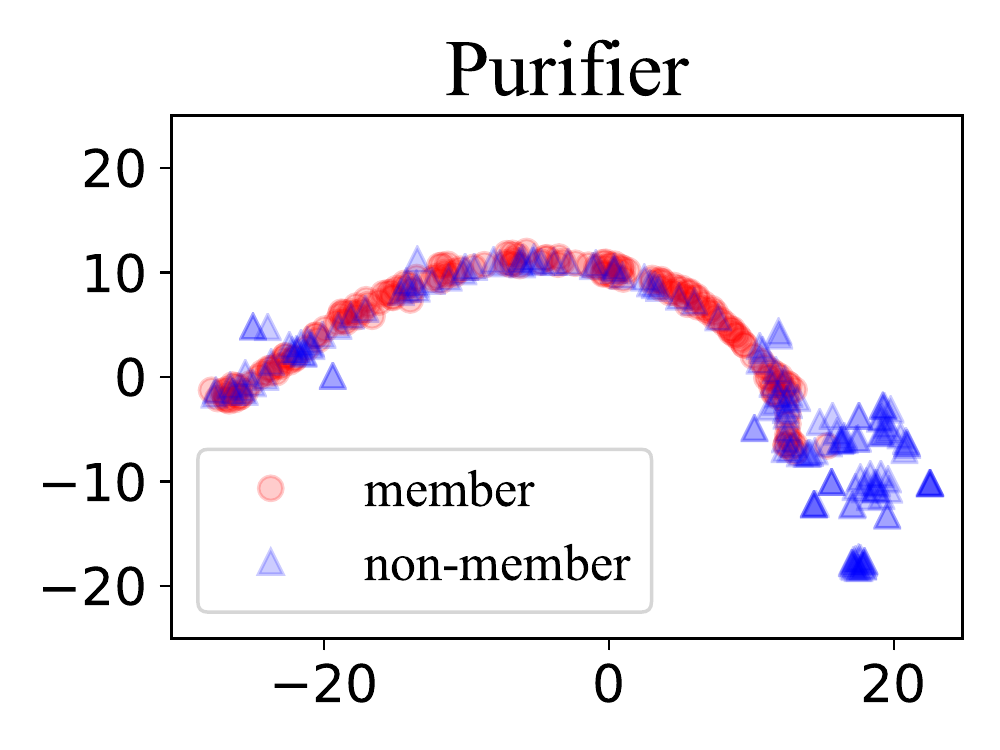}
		}
	\end{minipage}
	\caption{Confidence score vectors (predicted by a classifier in our experiment) projected into a 2D space. (a) No defense is used. (b) Purifier is used. After purification, the vectors are more tightly clustered with less dispersion.}
	\label{fig:intro_2dmap}
\end{figure}

In this paper, we propose a unified purification framework to defend data inference attacks by ``purifying'' the prediction scores. 
The framework takes the prediction scores of a trained target model as input and produces a purified version to satisfy one or both of these defense goals: (\rom{1}) preventing model inversion attack and (\rom{2}) preventing membership inference attack.
The intuition of the purification framework is to reduce the \textit{dispersion} of the confidence score vectors predicted by the target classifier no matter on members or non-members, which can be visualized in Figure~\ref{fig:intro_2dmap}. 
This brings two benefits. First, it is helpful to reduce the sensitivity of the prediction to the change of input data. In particular, changing the input data will lead to a smaller change of the confidence score vectors. 
It can decrease the correlation between the input data and the confidence scores. As a result, the model inversion against such classifier is inaccurate, and we achieve the Defense Goal \rom{1}. 
Second, it can lead to the reduction of the distinguishability of confidence score vectors between members and non-members. 
When the confidence score vectors become tightly clustered, they are less different from each other, which makes it more difficult for the attacker to distinguish them.
As a consequence, the membership inference is inaccurate, and we achieve the Defense Goal \rom{2}.
Note that, the purification framework is shown to incur negligible distortion to the original confidence scores while reducing their dispersion, which can preserve the useful information of the prediction.

We achieve the purification framework by training a \textit{purifier}, which takes the confidence scores of the target model as input and ``concentrate'' them on clear patterns.
It can become further specialized in the task of defending a particular attack through a specially designed adversarial learning process.
Specifically, we train an additional adversarial model for the Defense Goal \rom{1} (i.e., preventing model inversion), and train an additional discriminator for the Defense Goal \rom{2} (i.e., preventing membership inference).
The two additional models can be discarded after training. The purifier and the original classifier will work as a black box to classify the query data.

We design the purifier as an autoencoder~\cite{baldi_autoencoders_2012} which has been widely used to learn a latent representation of a dataset.
We train the purifier on a reference dataset, which is a common strategy in existing defenses to train their models~\cite{jia_memguard_2019, nasr_machine_2018}.
The reference set consists of non-members in terms of the target classifier.
The purifier is trained to minimize the reconstruction error of the original confidence scores predicted by the target classifier and the reconstructed version.
The autoencoder structure of the purifier is able to capture the latent patterns of the confidence scores on these non-members.
As a result, any confidence scores will be reconstructed around the learned latent patterns, as if they were predicted on non-members.
As we will show in the experiments, such reconstruction is effective in reducing the dispersion of the confidence scores while introducing insignificant distortion to them.

The purifier can be further specialized for a particular defense goal by an adversarial learning process.
Specifically, for the Defense Goal \rom{1}, we anticipate an additional adversarial model which adaptively performs model inversion attack against the purifier and the purifier keeps updating the prediction scores to minimize the inversion accuracy.
Eventually, the adversarial model evolves to be the supposedly strongest inversion model and the purifier purifies the prediction scores with minimized useful information for the mode inversion attack.
We formulate such training as a min-max game between the purifier and the adversarial model, and jointly train them.
For the Defense Goal \rom{2},  we adopt a discriminator when training the purifier to make the reconstruction more realistic, which is a similar training strategy as in the generative adversarial network~\cite{goodfellow_generative_2014}. 
The discriminator is trained to distinguish whether a confidence score is predicted on real data or a fake reconstruction, and the purifier learns to fool the discriminator to make mistakes.
Eventually, the purifier improves the realness of the reconstructed confidence scores.

The purifier can concurrently attain both defense goals when we jointly train the purifier, the adversarial model and the discriminator.
The result of the joint training is that, at the equilibrium point, 
the purifier can not only purify the confidence scores to reduce their dispersion but also become specialized in the task of defending each particular attack.

We extensively evaluate the purification framework on various benchmark datasets and model architectures which are widely used in previous work of data inferences. 
We empirically show that the membership inference attack and the model inversion attack, as two important data inference attacks, are related. Specifically, when the purifier is specialized in defending one attack, it naturally improves the defense performance against the other one.
Reducing the dispersion of the confidence score vectors can mitigate both attacks at the same time.
The defense performance against the model inversion attack significantly outperforms existing methods. For instance, our experimental results on the FaceScrub classifier~\cite{yang_neural_2019} (as shown in Figure~\ref{fig:facescrub}) visibly show that the inversion quality decreases substantially.
The purification framework is also effective in defending the membership inference attack. For instance, the inference accuracy decreases by up to 15\% in our experiments.
The utility loss is negligible, e.g., the classification accuracy loss is within 0.4\% and the confidence score distortion is less than 5.5\%.
Besides, the purifier is efficient. For example, its prediction time is 4,636 times less than one of the state-of-the-art defenses---MemGuard~\cite{jia_memguard_2019}.

\textbf{Contributions.} In summary, we make the following contributions in this paper.
\begin{itemize}
	
	\item We propose a unified purification framework to defend both model inversion attack and membership inference attack by reducing the dispersion of confidence score vectors.
	
	\item We empirically show the relation of both attacks. When the purifier is specialized in defending one attack, it naturally mitigates the other one.
	
	\item We extensively evaluate our approach and compare it with existing defenses on benchmark datasets.
	
\end{itemize}

\section{Inference Attacks on Machine Learning}

It has been shown that machine learning models are vulnerable to various inference attacks~\cite{shokri_membership_2017, yang_neural_2019, fredrikson_privacy_2014, yeom_privacy_2018}, which enables adversaries to get useful information about the target model from only the prediction APIs.
Depending on the inference goals, these inference attacks generally fall into two classes, i.e., \textit{model inference} and \textit{data inference}. Specifically, model inference aims at obtaining the information about the target model itself such as its parameters and architecture~\cite{tramer_stealing_2016, oh_towards_2018, wang_stealing_2018, orekondy_knockoff_2019}. Data inference, on the contrary, focuses on extracting information about the data on which the target model operates~\cite{shokri_membership_2017, yang_neural_2019, fredrikson_model_2015, yeom_privacy_2018, fredrikson_privacy_2014, wu_methodology_2016, ateniese_hacking_2015, wei_i_2018, salem_updates-leak_2020}. 
In this paper, we concentrate on two of the most important and exemplary data inference attacks, notably membership inference attack~\cite{shokri_membership_2017} and model inversion attack~\cite{yang_neural_2019, fredrikson_model_2015}.
In this section, we first introduce these two data inference attacks and then introduce existing defenses. Finally, we analyze the limitations of these methods when applied to both attacks.

\subsection{Data Inference Attacks on Machine Learning}
\label{sec:datainference}

Membership inference and model inversion attacks are two types of data inference attacks that threaten the security of machine learning. 
They differ in their inference goal.
\\\\
\noindent \textbf{Membership Inference Attack.}
In the membership inference attack, the attacker is asked to determine whether a given data record is part of the training data of the target model~\cite{shokri_membership_2017, nasr_comprehensive_2019, salem_ml-leaks_2018, jia_memguard_2019, long_towards_2017, long_understanding_2018}. 

\underline{\textit{Confidence-based Attack}}~\cite{shokri_membership_2017, salem_ml-leaks_2018}.
Shokri et al.~\cite{shokri_membership_2017} introduced membership inference against black-box models, where the attacker has access only to the prediction scores of the target model. 
To infer the membership, the attacker trains a binary classifier (also referred to as \textit{attack model}) which takes the confidence scores of the target model on a given data sample as input and predicts the data sample to be a member or non-member of the training dataset of the target model.
Prior to training the attack model, the attacker trains a set of \textit{shadow models} on an auxiliary dataset drawn from the same data distribution as the target model's training data to replicate the target model. 
The attack model is then trained on the confidence scores predicted by the shadow models instead of the target model on the members and non-members of the shadow models' training data.
Salem et al.~\cite{salem_ml-leaks_2018} further showed that it is sufficient to train only one shadow model to replicate the target model for membership inference attack. Besides, they also showed that ranking the elements in the confidence score vectors before inputting them to the attack model could improve the inference accuracy. For example, their experimental results show that only the top one or three highest values in the confidence vector are sufficient to result in effective membership inference.

\underline{\textit{Label-based Attack}}~\cite{yeom_privacy_2018}. Yeom et al.~\cite{yeom_privacy_2018} proposed a label-based method which predicts that an input sample is a member of the target classifier if and only if the classifier gives the correct label on the sample. This method is simple and can be applied when the adversary is given only the label. 

\underline{\textit{Confidence \& Label-based Attack}}~\cite{nasr_machine_2018}.
Nasr et al.~\cite{nasr_machine_2018} extended the attack model by also taking the label information as input. Their attack model is composed of three neural networks. The first two networks operate on the confidence score vector and the one-hot encoded label respectively. They have the same size of input dimension, i.e., the number of classes of the target model. The third network operates on the concatenation of the output of the first two networks and predicts the membership. This method assumes that the attacker has a subset of the members and non-members of the target model's training data, and therefore, no shadow model is trained.

More settings of membership inference attack have been explored in the literature such as while-box membership inference attacks~\cite{nasr_comprehensive_2019, leino_stolen_2020}.
There are also membership inference attacks in federated learning~\cite{nasr_comprehensive_2019, melis_exploiting_2019} and against generative models~\cite{hayes2019logan}. In this paper, we consider membership inference attack in the black-box setting against standalone centralized classification models.
\\\\
\noindent \textbf{Model Inversion Attack.} 
We consider the black-box model inversion setting~\cite{yang_neural_2019} where the attacker wants to reconstruct an input sample from its confidence score vector predicted by the target model.
Fredrikson et al.~\cite{fredrikson_model_2015} proposed the initial model inversion attack in the white-box setting. 
However, this attack produces only a representative sample of a training class instead of reconstructing a specific input sample from the confidence score vector.

Yang et al.~\cite{yang_neural_2019} proposed a model inversion attack that is able to reconstruct specific input samples in the black-box setting. Specifically, they train a separate inversion model on an auxiliary dataset which acts as the inverse of the target model. 
The inversion model takes the confidence scores of the target model as input and tries to reconstruct the original input data. 
Their experimental results showed substantial improvement of the inversion accuracy over previous works.

Some other methods were proposed to infer the sensitive attributes~\cite{fredrikson_privacy_2014, wu_methodology_2016, hidano_model_2017, zhang_secret_2019} or statistical information~\cite{ateniese_hacking_2015} about the training data.
There are also inversion attacks~\cite{hitaj_deep_2017} in federated learning where the attacker has white-box access to the global model.
In this paper, we focus on model inversion attack that aims to reconstruct the input data in the black-box setting against standalone centralized classification models.

\subsection{Defenses against Data Inference Attacks}
\label{sec:existing_defense}

Previous approaches mostly focus on defending the membership inference attack. 
However, little has been studied about the defense of the model inversion attack on classification models.
Therefore, we introduce existing defenses against the membership inference attack as typical examples in defending data inference attacks in the literature.

A number of studies make use of various regularization techniques and ensemble learning to reduce overfitting as a defense against membership inference attack. When a model overfits on training data (i.e., members), it behaves more confidently on their training data than others. 
As a result, 
the confidence scores of the model on members and non-members present different patterns, which enables the attacker to distinguish them. We summarize the typical defense methods that reduce overfitting in the following.

	\underline{\textit{$L_2$-Regularizer}}~\cite{shokri_membership_2017}. The $L_2$ regularizer is an $L_2$ norm of the model parameters added as a weighted penalty term  to the original loss function. In \cite{shokri_membership_2017}, the authors showed that using $L_2$-regularizer to train the target model can help mitigate membership inference attack.
	
	 \underline{\textit{Dropout}}~\cite{salem_ml-leaks_2018}. Dropout is another technique used to regularize neural networks~\cite{srivastava_dropout_2014}. It works by dropping a neuron with a certain probability during the network training. It can be used to mitigate the membership inference attack~\cite{salem_ml-leaks_2018}.
	
	 \underline{\textit{Min-Max Game}}~\cite{nasr_machine_2018}. Nasr et al.~\cite{nasr_machine_2018} proposed to add an adversarial regularizer to the loss function of the target model such that it is trained to minimize the prediction loss and also to maximize the membership privacy. The training process is formulated as a min-max optimization problem.
	
	 \underline{\textit{Model Stacking}}~\cite{salem_ml-leaks_2018}. Model stacking is essentially an ensemble approach which combines multiple simple classifiers as a complicated one to make the final prediction~\cite{opitz_popular_1999, polikar_ensemble_2006}. It is often used as a way of reducing overfitting~\cite{sollich1996learning}, and can be leveraged to mitigate membership inference attack~\cite{salem_ml-leaks_2018}.

While most existing defenses focus on reducing overfitting to mitigate membership inference attack, there are also approaches proposed from different angles. We summarize them in the following.

	 \underline{\textit{MemGuard}}~\cite{jia_memguard_2019}. Jia et al.~\cite{jia_memguard_2019} studied to transform the confidence score vector into an adversarial example to evade the membership classification of the attack model. Specifically, the defender adds carefully-crafted noise to the confidence score vector predicted by the target model so as to turn it into an adversarial example. To this end, the defender first trains its own ``attack model'' which works similarly as the attacker's attack model, and thus it can craft the adversarial example against its attack model in a white-box manner. Such adversarial example is also likely to evade the membership classification of the attacker's attack model due to the transferability of adversarial examples~\cite{demontis_why_2019, papernot2016transferability, papernot_practical_2017}.

	 \underline{\textit{Differential Privacy.}} Differential privacy~\cite{dwork_calibrating_2006} is a widely used privacy-preserving technique. 
	Some methods use differential privacy to add noise to the objective function of the model~\cite{chaudhuri2011differentially, iyengar_towards_2019, kifer_private_2012}, while others add noise to the gradient of the model during minimizing the objective function~\cite{abadi_deep_2016, bassily_private_2014, song_stochastic_2013, wang_differentially_2017, yu_differentially_2019}. Differential privacy is able to provide theoretical privacy guarantee but incurs significant loss of classification accuracy~\cite{nasr_machine_2018, li_membership_2020}.

\subsection{Limitations When Applied to Defending Both Attacks}

Most of the previous defenses rely on reducing overfitting to defend membership inference attack.
However,
it is shown that overfitting is not the only reason that causes membership inference attack~\cite{shokri_membership_2017}. Even if different machine learning models are overfitted to the same degree, they could leak different amounts of membership information. Specifically, due to their different structures, they might ``remember'' different amounts of information about their training data.
Actually, the attacker exploits the information about how the target model's confidence scores distinguish members from non-members to launch membership inference attack~\cite{shokri_membership_2017}.
Our experiments show that our approach can mitigate membership inference without changing the generalization ability of the classifier.

MemGuard~\cite{jia_memguard_2019}, instead of reducing overfitting, turns the confidence score vector into an adversarial example to fool the attacker's membership classifier. Its effectiveness relies on the transferability~\cite{papernot2016transferability} of adversarial examples, which does not generally reduce the distinguishability of confidence scores as well. Moreover, it incurs much more overhead in the prediction time
than other methods. In our experiments, the prediction time of MemGuard is 4,636 times more than our approach.

Differential privacy, though provides a theoretical guarantee of privacy, has a significant impact on the classification accuracy of the target model compared with other defense mechanisms~\cite{nasr_machine_2018}. 
Besides, although differential privacy prevents an attacker from gaining additional information by including or excluding an individual data record, 
the information leakage from the released prediction scores (through which an attacker can perform model inversion attack) remains unclear.

As we will show in the experiments, the membership inference attack and the model inversion attack are actually related. 
For example, reducing the dispersion of the confidence score vectors can mitigate them at the same time. 
However, we show that previous approaches that defend the membership inference attack have very limited defense effect on the model inversion attack.
To the best of our knowledge, no known method of defending both attacks is available.

\section{Problem Formulation}

We focus on the supervised learning, more specifically, on training classification models (classifiers) using neural networks~\cite{lecun2015deep}.
There are three parties in our problem, namely \textit{model owner}, \textit{attacker} and \textit{defender}. The model owner trains a machine learning classifier on its training dataset. We refer to this classifier as \textit{target classifier}. 
The attacker aims to launch data inference attacks against the target classifier. 
The defender aims to provide protection to the target classifier against the attacker.

\subsection{Model Owner}

The model owner trains a machine learning classifier $F$ on its training dataset $D_{train}$. It also has a validation dataset $D_{val}$ to test whether $F$ functions properly. Both $D_{train}$ and $D_{val}$ are drawn from the same underlying data distribution $p_x(\vec{x})$.
The classifier $F$ is trained with the goal of making predictions on unseen data which we refer to as test dataset $D_{test}$.
Let $\vec{x}$ represent the data drawn from $p_x$, and $\vec{y}$ be the vectorized class of $\vec{x}$. The training objective is to find a function $F$ to well approximate the relation between each data point $(\vec{x}, \vec{y})$. Formally, we have $F: \vec{x}\mapsto\vec{y}$.
The training process is to optimize an objective function $L(F)$. It terminates typically when the classification accuracy on the validation set $D_{val}$ achieves the best~\cite{caruana2001overfitting}. The model owner releases the trained classifier $F$ as a black box, for example, as a cloud service, and provides prediction APIs to users. The users can query $F$ with their own data sample $\vec{x} \in D_{test}$ through the prediction APIs. 
The classifier $F$ returns a prediction score vector $F(\vec{x})$ to the users. The prediction score vector is a probability distribution of the classifier's confidence over all the possible classes. For example, the $i$-th element $F(\vec{x})_i$ is the probability of the data $\vec{x}$ belonging to class $i$. We usually take the class with the maximum probability to be the predicted label of the data $\vec{x}$.

\subsection{Attacker}

The attacker aims at performing data inference attacks against the target classifier $F$. We consider that the classifier $F$ works as a black-box ``oracle'' to the attacker, i.e., the attacker can only query $F$ with its data sample $\vec{x}$ and obtain the prediction scores $F(\vec{x})$.
The attacker is also assumed to have auxiliary information $\mathcal{A}$ such as the ground-truth label of $\vec{x}$ and a set of data samples drawn from a similar data distribution as the target classifier's training data distribution. Formally, given a prediction vector $F(\vec{x})$ on some victim data point $\vec{x}$, the attacker wants to find an attack function $A(F(\vec{x}), \mathcal{O}(F), \mathcal{A})$ in the following:
\begin{equation}
\small
	A(F(\vec{x}), \mathcal{O}(F), \mathcal{A}) = 
	\begin{cases}
	m \in \{0, 1\},	& \text{membership inference} \\ 
	\tilde{\vec{x}},		& \text{model inversion}
	\end{cases}
\end{equation}
where $\mathcal{O}(F)$ represents the attacker's black-box access to the oracle classifier $F$, $m$ denotes the membership of the victim data $\vec{x}$, and $\tilde{\vec{x}}$ is the reconstruction of $\vec{x}$ . The membership $m$ has two possible values: $0$ and $1$, where $m=1$ indicates that $\vec{x}$ is a member of the target classifier's training data while $m=0$ means non-member.

\subsection{Defender}
\label{sec:defender}

The defender protects the target classifier from data inference attacks.
It could be the model owner or a trusted third party who has access to the target classifier's prediction score vectors and a reference dataset $D_{ref}$ as previous approaches~\cite{jia_memguard_2019, nasr_machine_2018} assume.
For any query to the target classifier from users, the defender modifies the prediction score vector of the target classifier to prevent the attacker from performing accurate data inference attacks before returning it to users. The attacker has access only to the modified prediction scores from the defender.
In particular, the defender wants to achieve the following goals:
\begin{itemize}
	
	\item \textbf{Defense Goal.} The defender aims to achieve one or both of these defense goals. 
	(\textbf{Defense Goal \rom{1}}) The defender wants to make the error of $A$ on reconstructing the input data $\vec{x}$ large enough, such that the attacker is unable to infer anything specific about $\vec{x}$ from the $\tilde{\vec{x}}$.
	(\textbf{Defense Goal \rom{2}}) The defender wants to make the attack function $A$ inaccurate at inferring the membership of a given data sample.
	
	\item \textbf{Utility Goal.} The defender aims to introduce negligible classification accuracy loss as well as insignificant confidence score distortion.
	
	\item \textbf{Efficiency Goal.} The defense mechanism should incur acceptable overhead in training time and testing time.
	
\end{itemize}

\section{Approach: Purification Framework}

\def\layersep{1.5cm}
\def\modelsep{2cm}
\def\nodesep{1cm}

\begin{figure*}  
	\centering  
	
	\subfigure[Purifier \rom{1}]  
	{  	
		\begin{tikzpicture}[-,draw=black, node distance=\layersep,transform shape,rotate=0,scale=0.8]  
		\label{fig: model_inversion}
		
		\node[draw, dashed, drop shadow, fill=yellow!25, minimum width=3.2*\nodesep,minimum height=2.6*\layersep,rounded corners] (release) at (0,1.2*\nodesep) {};
		
		\node[rotate=0] (input) at (0.2*\nodesep, -0.9*\nodesep) {$\vec{x}$};
		\node[draw,thick,aspect=0.2,cylinder,drop shadow,fill=white,minimum width=\nodesep, minimum height=0.7*\nodesep, rotate=90] (input) at (0, -1.5*\nodesep) {\small \rotatebox{270}{ $D_{ref}$}};
		
		\node[draw,thick,fill=gray!20,minimum width=3*\nodesep,minimum height=0.7*\layersep,rounded corners] (classifier) at (0,0) {};
		\node[rotate=0] (b) at (0, 0) {Target Classifier $F$};
		
		\path[->,thick,line width=1pt,rotate=90] (input) edge (classifier);
		
		\node[draw,fill=white,minimum width=3*\nodesep,minimum height=0.3*\layersep,rounded corners] (prediction) at (0,1.2*\nodesep) {};
		\node[rotate=0] (b) at (0, 1.2*\nodesep) {$F(\vec{x})$};
		
		\path[->,thick,line width=1pt,rotate=90] (classifier) edge (prediction);
		
		\node[draw,thick,fill=gray!50,minimum width=3*\nodesep,minimum height=0.7*\layersep,rounded corners] (purifier) at (0,2.4*\nodesep) {};
		\node[rotate=0] (b) at (0, 2.4*\nodesep) {\makecell[c]{Purifier $G$ \\\textcolor{blue}{ \scriptsize $\min \mathbb{E}[L_1 + \lambda L_2 + \alpha L_3]$}   }   };
		
		\path[->,thick,line width=1pt,rotate=90] (prediction) edge (purifier);
		
		\node[draw,fill=white,minimum width=3*\nodesep,minimum height=0.3*\layersep,rounded corners] (purified prediction) at (0,3.6*\nodesep) {};
		\node[rotate=0] (b) at (0, 3.6*\nodesep) {$G(F(\vec{x}))$};
		
		\path[->,thick,line width=1pt,rotate=90] (purifier) edge (purified prediction);
		
		\node[draw,drop shadow, thick,fill=gray!50,minimum width=3*\nodesep,minimum height=0.7*\layersep,rounded corners] (adversarial model) at (0,4.8*\nodesep) {};
		\node[rotate=0] (b) at (0, 4.8*\nodesep) {\makecell[c]{\small Adversarial Model $H$\\\textcolor{blue}{\small $\max \mathbb{E}[L_3]$}}  };
		
		\path[->,thick,line width=1pt,rotate=90] (purified prediction) edge (adversarial model);
		
		\node[draw,fill=white,minimum width=3*\nodesep,minimum height=0.3*\layersep,rounded corners] (reconstructed input) at (0,6*\nodesep) {};
		\node[rotate=0] (b) at (0, 6*\nodesep) { $H(G(F(\vec{x})))$};
		
		\path[->,thick,line width=1pt,rotate=90] (adversarial model) edge (reconstructed input);
		
		\node[draw, fill=gray!10,minimum height=0.5cm,minimum width=2.5cm,rounded corners] (loss12) at (-3.5*\nodesep,4.8*\nodesep) {\scriptsize $L_3 = -\mathcal{R}(x, H(G(F(\vec{x}))))$};
		\path[->,dotted,line width=0.7pt,in=-110, out=180] (input.north) edge node [fill=white] { $\vec{x}$}  (loss12.south);
		\path[->,dotted, line width=0.7pt,in=60, out=190] (reconstructed input.west) edge (loss12.north);
		
		\node[draw, fill=gray!10,minimum height=0.5cm,minimum width=2.5cm,rounded corners] (loss4) at (-3.5*\nodesep,2.4*\nodesep) {\makecell[c]{\scriptsize $L_1 = \mathcal{R}(G(F(\vec{x})), F(\vec{x}))$\\ \tiny $L_2 = \mathcal{L}(G(F(\vec{x})), \argmax F(\vec{x}))$ }  };
		\path[->,dotted, line width=0.7pt,in=-70, out=170] (prediction.west) edge  (loss4.south);
		\path[->,dotted,line width=0.7pt,in=70, out=190] (purified prediction.west) edge (loss4.north);
		
		\end{tikzpicture}
	}  
	\qquad
	\subfigure[Purifier \rom{2}]  
	{  
		
		\begin{tikzpicture}[-,draw=black, node distance=\layersep,transform shape,rotate=0,scale=0.8]  
		\label{fig: membership_inference}
		\node[draw, dashed, drop shadow, fill=yellow!25, minimum width=3.2*\nodesep,minimum height=2.6*\layersep,rounded corners] (release) at (0,1.2*\nodesep) {};
		
		\node[rotate=0] (input) at (0.2*\nodesep, -0.9*\nodesep) {$\vec{x}$};
		\node[draw,thick,aspect=0.2,cylinder,drop shadow,fill=white,minimum width=\nodesep, minimum height=0.7*\nodesep, rotate=90] (input) at (0, -1.5*\nodesep) {\small \rotatebox{270}{ $D_{ref}$}};
		
		\node[draw,thick,fill=gray!20,minimum width=3*\nodesep,minimum height=0.7*\layersep,rounded corners] (classifier) at (0,0) {};
		\node[rotate=0] (b) at (0, 0) {Target Classifier $F$};
		
		\path[->,thick,line width=1pt,rotate=90] (input) edge (classifier);
		
		\node[draw,fill=white,minimum width=3*\nodesep,minimum height=0.3*\layersep,rounded corners] (prediction) at (0,1.2*\nodesep) {};
		\node[rotate=0] (b) at (0, 1.2*\nodesep) {$F(\vec{x})$};
		
		\path[->,thick,line width=1pt,rotate=90] (classifier) edge (prediction);
		
		\node[draw,thick,fill=gray!50,minimum width=3*\nodesep,minimum height=0.7*\layersep,rounded corners] (purifier) at (0,2.4*\nodesep) {};
		\node[rotate=0] (b) at (0, 2.4*\nodesep) {\makecell[c]{Purifier $G$ \\\textcolor{blue}{\scriptsize $\min \mathbb{E}[L_1 + \lambda L_2 + \beta L_4]$}   }    };
		
		\path[->,thick,line width=1pt,rotate=90] (prediction) edge (purifier);
		
		\node[draw,fill=white,minimum width=3*\nodesep,minimum height=0.3*\layersep,rounded corners] (purified prediction) at (0,3.6*\nodesep) {};
		\node[rotate=0] (b) at (0, 3.6*\nodesep) {$G(F(\vec{x}))$};
		
		\path[->,thick,line width=1pt,rotate=90] (purifier) edge (purified prediction);

		\node[draw,drop shadow, thick,fill=gray!50,minimum width=3*\nodesep,minimum height=0.7*\layersep,rounded corners] (discriminator) at (0,4.8*\nodesep) {};
		\node[rotate=0] (b) at (0, 4.8*\nodesep) {\makecell[c]{Discriminator $I$\\\textcolor{blue}{\small $\max \mathbb{E}[L_5]$}  }};
		
		\path[->,thick,line width=1pt,rotate=90] (purified prediction) edge (adversarial model);
		
		\path[->,thick,line width=1pt,rotate=90] (purified prediction) edge (discriminator);
		
		\node[draw,fill=white,minimum width=3*\nodesep,minimum height=0.3*\layersep,rounded corners] (discrminator output) at (0, 6*\nodesep) {};
		\node[rotate=0] (b) at (0, 6*\nodesep) { $I(G(F(\vec{x})))$};
		
		\path[->,thick,line width=1pt,rotate=90] (discriminator) edge (discrminator output);
		
		\node[draw,fill=gray!10,minimum height=0.5cm,minimum width=2.5cm,rounded corners] (loss35) at (3.5*\nodesep,2.4*\nodesep) {\makecell[c]{\scriptsize $L_4 = \log(1-I(G(F(\vec{x}))))$\\ \scriptsize $L_5 = \log(I(F(\vec{x}))) + L_4$ }  };
		\path[->,dotted, line width=0.7pt,in=-110, out=10] (prediction.east) edge  (loss35.south);
		\path[->,dotted,line width=0.7pt,in=110, out=-10] (purified prediction.east) edge (loss35.north);

		\end{tikzpicture}
		
	}
	\subfigure[Purifier \rom{1} \& \rom{2}]{
		\begin{tikzpicture}[-,draw=black, node distance=\layersep,transform shape,rotate=0,scale=0.8]  
		\label{fig: both_attacks}
		\node[draw, dashed, drop shadow, fill=yellow!25, minimum width=4*\nodesep,minimum height=2.6*\layersep,rounded corners] (release) at (0,1.2*\nodesep) {};
		
		\node[rotate=0] (input) at (0.2*\nodesep, -0.9*\nodesep) {$\vec{x}$};
		\node[draw,thick,aspect=0.2,cylinder,drop shadow,fill=white,minimum width=\nodesep, minimum height=0.7*\nodesep, rotate=90] (input) at (0, -1.5*\nodesep) {\small \rotatebox{270}{ $D_{ref}$}};
		
		\node[draw,thick,fill=gray!20,minimum width=3.8*\nodesep,minimum height=0.7*\layersep,rounded corners] (classifier) at (0,0) {};
		\node[rotate=0] (b) at (0, 0) {Target Classifier $F$};
		
		\path[->,thick,line width=1pt,rotate=90] (input) edge (classifier);
		
		\node[draw,fill=white,minimum width=3.8*\nodesep,minimum height=0.3*\layersep,rounded corners] (prediction) at (0,1.2*\nodesep) {};
		\node[rotate=0] (b) at (0, 1.2*\nodesep) {$F(\vec{x})$};
		
		\path[->,thick,line width=1pt,rotate=90] (classifier) edge (prediction);
		
		\node[draw,thick,fill=gray!50,minimum width=3.8*\nodesep,minimum height=0.7*\layersep,rounded corners] (purifier) at (0,2.4*\nodesep) {};
		\node[rotate=0] (b) at (0, 2.4*\nodesep) {\makecell[c]{Purifier $G$ \\\textcolor{blue}{\scriptsize $\min \mathbb{E}[L_1 + \lambda L_2 + \alpha L_3 + \beta L_4]$}   }    };

		\path[->,thick,line width=1pt,rotate=90] (prediction) edge (purifier);
		
		\node[draw,fill=white,minimum width=3*\nodesep,minimum height=0.3*\layersep,rounded corners] (purified prediction) at (0,3.6*\nodesep) {};
		\node[rotate=0] (b) at (0, 3.6*\nodesep) {$G(F(\vec{x}))$};
		
		\path[->,thick,line width=1pt,rotate=90] (purifier) edge (purified prediction);
		
		\node[draw,drop shadow, thick,fill=gray!50,minimum width=3*\nodesep,minimum height=0.7*\layersep,rounded corners] (adversarial model) at (-2*\nodesep,4.8*\nodesep) {};
		\node[rotate=0] (b) at (-2*\nodesep, 4.8*\nodesep) {\makecell[c]{\small Adversarial Model $H$\\\textcolor{blue}{\small $\max \mathbb{E}[L_3]$}}  };
		
		\node[draw,drop shadow, thick,fill=gray!50,minimum width=3*\nodesep,minimum height=0.7*\layersep,rounded corners] (discriminator) at (2*\nodesep,4.8*\nodesep) {};
		\node[rotate=0] (b) at (2*\nodesep, 4.8*\nodesep) {\makecell[c]{Discriminator $I$\\\textcolor{blue}{\small $\max \mathbb{E}[L_5]$}  }};
		
		\path[->,thick,line width=1pt,rotate=90] (purified prediction) edge (adversarial model);
		
		\path[->,thick,line width=1pt,rotate=90] (purified prediction) edge (discriminator);
		
		\node[draw,fill=white,minimum width=3*\nodesep,minimum height=0.3*\layersep,rounded corners] (reconstructed input) at (-2*\nodesep,6*\nodesep) {};
		\node[rotate=0] (b) at (-2*\nodesep, 6*\nodesep) { $H(G(F(\vec{x})))$};
		
		\path[->,thick,line width=1pt,rotate=90] (adversarial model) edge (reconstructed input);
		
		\node[draw,fill=white,minimum width=3*\nodesep,minimum height=0.3*\layersep,rounded corners] (discrminator output) at (2*\nodesep,6*\nodesep) {};
		\node[rotate=0] (b) at (2*\nodesep, 6*\nodesep) { $I(G(F(\vec{x})))$};
		
		\path[->,thick,line width=1pt,rotate=90] (discriminator) edge (discrminator output);

		\end{tikzpicture}
	}
	
	\caption{
		Architecture of the purification framework for different defense goals.
		The base common component of the purification frameworks is the purifier $G$. It has a similar structure as autoencoder as shown in Figure~\ref{fig:autoencoder} and thus can be used to learn the distribution of the confidence score vector $F(\vec{x})$.
		The target classifier $F$ is fixed during training $G$. After training, $G$ and $F$ will work as a black box (i.e., the yellow part) to classify query data.
		(a) The purifier specialized in defending model inversion attack.
		(b) The purifier specialized in defending membership inference attack.
		(c) The purifier specialized in defending both attacks.
	}
	\label{fig:architecture}
\end{figure*}
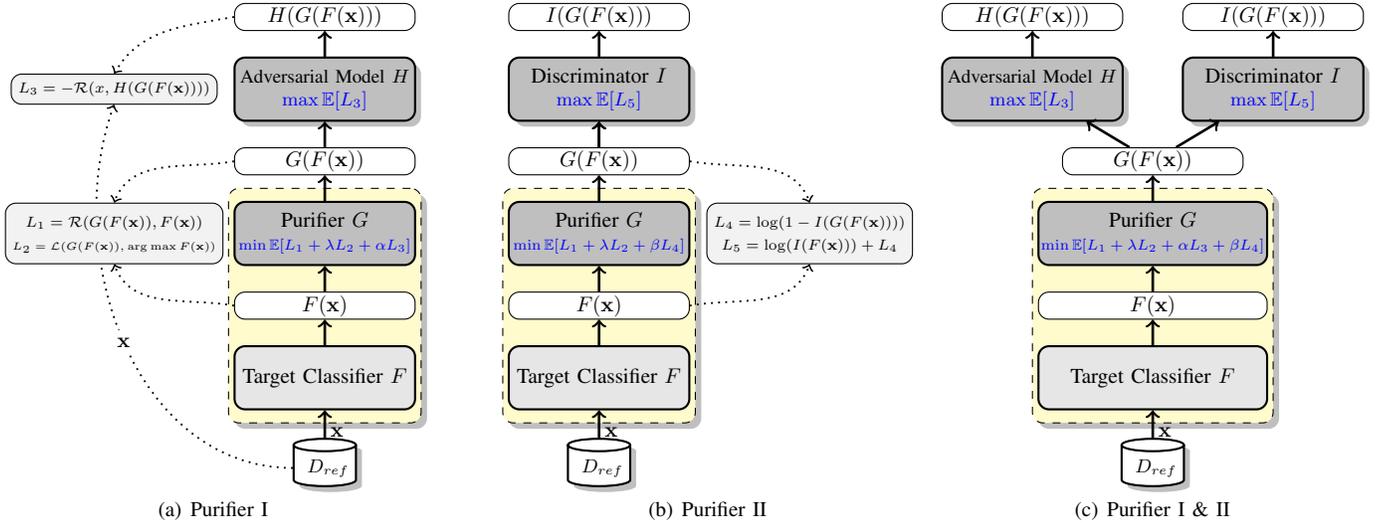

We propose a purification framework to defend data inference attacks, i.e., model inversion attack and membership inference attack, by purifying the confidence scores of the target classifier such that the attacker could not exploit the purified confidence scores to infer inversion and membership information about the data.
The purification framework does not tamper with the training process of the target classifier.

The architecture of the purification framework for each defense goal is shown in Figure~\ref{fig:architecture}.
The common component of the purification framework is a \textit{purifier} $G$ for all defense goals. 
It takes the confidence score vectors of the target classifier $F$ as input and reshapes it to reduce their dispersion as well as to preserve the utility of the classifier. 
On the one hand, the reduction of the dispersion can make the confidence score vectors less sensitive to the change of the input data, which decreases their correlation and results in the mitigation of model inversion attack.
On the other hand, the confidence score vectors become more tightly clustered, and thus become more similar with each other. This increases the attacker's difficulty to distinguish them and mitigates the membership inference attack.
The purifier $G$ can be further specialized in the task of defending each particular attack in an adversarial environment. To this end, we train an additional adversarial model $H$ for the Defense Goal \rom{1} and an additional discriminator $I$ for the Defense Goal \rom{2}. 
After training, $H$ and $I$ can be discarded. The purifier $G$ and the target classifier $F$ will work as a black box to classify the query data.

\subsection{Base of Purification: Purifier $G$}

\begin{figure}[t]
	\begin{center}
		\includegraphics[width=0.7\linewidth]{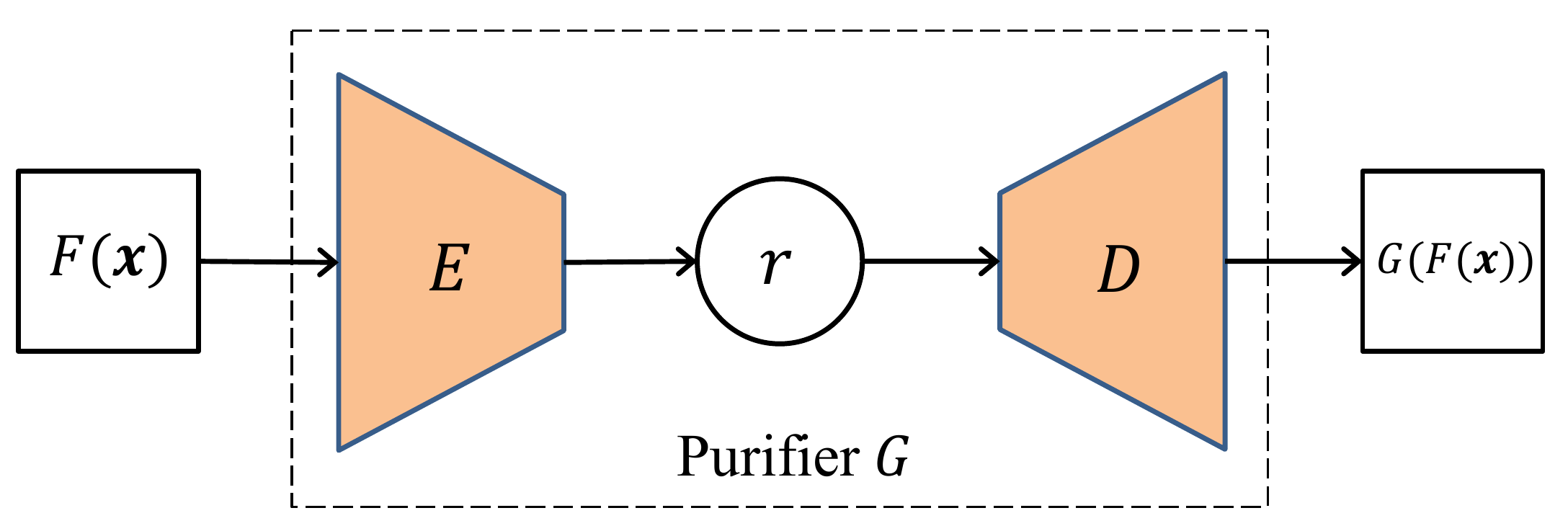}
		\caption{Architecture of purifier $G$. It consists of an encoder $E$ which maps the confidence score vector $F(\vec{x})$ predicted by $F$ to a latent representation $r$ and a decoder $D$ which maps $r$ to a reconstruction $G(F(\vec{x}))$.}
		\label{fig:autoencoder}
	\end{center}
\end{figure}

The purifier $G$ is used to reduce the dispersion of the confidence scores as well as to preserve the utility of the classifier.
We design $G$ as an autoencoder~\cite{baldi_autoencoders_2012} as shown in Figure~\ref{fig:autoencoder}.
We train it on the confidence scores predicted by $F$ on the defender's reference dataset $D_{ref}$, which consists of non-member data. 
The purifier is optimized to reconstruct the input confidence score vectors on $D_{ref}$, through which it learns their latent patterns.
After training, the purifier is able to reshape any confidence scores to concentrate on the learned latent patterns, as if they were predicted on non-member data. 
In order to preserve the classification accuracy, we train $G$ to also produce the label predicted by $F$ by adding a cross entropy loss function. Formally, $G$ is trained to minimize the following objective function.
\begin{equation}
\begin{split}
L(G) = \underset{\vec{x} \sim p_{r}(\vec{x})}{\mathbb{E}} &[\mathcal{R}(G(F(\vec{x})), F(\vec{x})) 
\\&+ \lambda \mathcal{L}(G(F(\vec{x}), \argmax F(\vec{x})))]
\end{split}
\end{equation}
where $p_r(\vec{x})$ represents the conditional probability of $\vec{x}$ for samples in $D_{ref}$, $\mathcal{R}$ is a reconstruction loss function (we use $L_2$ norm) and $\mathcal{L}$ is the cross entropy loss function. The parameter $\lambda$ controls the balance of the two loss functions during optimization.

\subsection{Specialized $G$ for Model Inversion Attack}

The defense performance of $G$ against the model inversion attack can be further improved when we jointly train $G$ and the adversarial model $H$ in an adversarial process.
Figure~\ref{fig: model_inversion} presents the architecture of the purification framework (referred to as Purifier \rom{1}) in this case.
In the model inversion attack, the attack function $A$ could be constructed in any unexpected ways with the goal of bypassing a particular defense mechanism. 
In response to this, we anticipate a supposedly strongest inversion function $H$ and train $G$ to minimize the inversion accuracy of $H$. Intuitively, the attacker could not design a better attack function $A$ than $H$ which is already prepared during the training of $G$. 
We model such training process as a min-max adversarial game between $G$ and $H$ which is a similar way as many adversarial processes for machine learning~\cite{dumoulin_adversarially_2017, goodfellow_generative_2014, miyato_virtual_2019, nasr_machine_2018}. 
Formally, $H$ is trained to optimize the following training objective.
\begin{equation}
\label{eq:loss_h}
\max_{H}\underset{\vec{x} \sim p_r(\vec{x})}{\mathbb{E}} [-\mathcal{R} (\vec{x}, H(G(F(\vec{x}))))]
\end{equation}
where $\mathcal{R}$ is the reconstruction loss function (we use $L_2$ norm).
This is exactly the same training objective of the attacker's inversion model in the model inversion attack~\cite{yang_neural_2019}.
The purifier $G$ learns to minimize the inversion accuracy by optimizing the following objective function.
\begin{equation}
\label{eq:loss_g}
\min_{G}\underset{\vec{x} \sim p_{r}(\vec{x})}{\mathbb{E}} [L(G) - \alpha \mathcal{R} (\vec{x}, H(G(F(\vec{x}))))]
\end{equation}
where $\alpha$ is a parameter that controls the trade-off between $L(G)$ and $\mathcal{R}$.
This training objective encourages $G$ to not only learn the task indicated by $L(G)$ but also maximize the inversion error of $H$. As a result, $G$ becomes more robust against the model inversion attack.

The two objective functions (i.e., function \ref{eq:loss_h} and \ref{eq:loss_g}) need to be solved jointly to find the equilibrium point. Therefore, we formalize them in one min-max optimization problem.
\begin{align}
\label{eq:min_max_loss}
\min_{G} \max_{H} \underset{\vec{x} \sim p_{r}(\vec{x})}{\mathbb{E}} [L(G) - \alpha \mathcal{R} (\vec{x}, H(G(F(\vec{x}))))]
\end{align}
The purifier $G$ and the adversarial model $H$ are trained alternatively to find the best responses against each other in one mini-batch. 

\subsection{Specialized $G$ for Membership Inference Attack}

The purifier $G$ can be further specialized in the task of defending the membership inference attack when we train it using the adversarial loss from the discriminator $I$.
Figure~\ref{fig: membership_inference} shows the architecture of the purification framework (referred to as Purifier \rom{2}) in this case.
The original $G$ is trained to reconstruct any confidence vectors to concentrate on the learned latent patterns, as if they were predicted on non-members. This helps reduce the distinguishability of the confidence scores between members and non-members because their dispersion decreases.
Hence, it increases the difficulty for the attacker to distinguish their membership.
However,
the purified confidence scores can be further improved to evade the membership classification if they behave like they are predicted on the real data rather than the fake reconstruction.
To this end, we use an additional discriminator $I$ to aid the training of $G$ in an adversarial manner. Specifically, $I$ is trained to distinguish real confidence scores and fake reconstructions by optimizing the following objective function.
\begin{equation}
\label{eq:i_loss}
	\max_{I}\underset{\vec{x} \sim p_r(\vec{x})}{\mathbb{E}} [\log I(F(\vec{x})) + \log(1 - I(G(F(\vec{x}))))]
\end{equation}

The purifier $G$ is trained to fool $I$ to misclassify the reconstructed confidence scores $G(F(\vec{x}))$ as real ones by optimizing the following objective function.
\begin{equation}
\label{eq:g_i_loss}
	\min_{G} \underset{\vec{x} \sim p_{r}(\vec{x})}{\mathbb{E}} [L(G) + \beta \log(1 - I(G(F(\vec{x}))))]
\end{equation}
where $\beta$ is a parameter controlling the importance of the loss functions during training.

We formalize the two objective functions (i.e., function \ref{eq:i_loss} and \ref{eq:g_i_loss}) in one min-max optimization problem to jointly train $G$ and $I$ to find the equilibrium point.
\begin{equation}
\label{eq:min_max_g_i}
\begin{split}
\min_{G} \max_{I}  \underset{\vec{x} \sim p_{r}(\vec{x})}{\mathbb{E}} [L(G) &+ \beta \log I(F(\vec{x})) \\
&+ \beta( \log(1 - I(G(F(\vec{x})))))]
\end{split}
\end{equation}
Similarly, the purifier $G$ and the discriminator $I$ are trained alternatively in one mini-batch to find the best responses against each other. 
Note that, the discriminator $I$ is used to distinguish real confidence scores or fake reconstructions, which helps refine the purified confidence scores to become as realistic as possible.
This is different from the method of Min-Max Game~\cite{nasr_machine_2018} where the discriminator (also referred to as inference model~\cite{nasr_machine_2018}) is used as a membership classifier and joins the training of the target classifier.

\subsection{Specialized $G$ for Both Attacks}

The purifier $G$ can be specialized in the tasks of defending both attacks if we jointly train $G$, $H$ and $I$. Figure~\ref{fig: both_attacks} presents the architecture of the purification framework (referred to as Purifier \rom{1} \& \rom{2}).
We formalize the joint training in a min-max-max optimization problem.
\begin{equation}
\small
\begin{split}
\min_{G} \max_{H} \max_{I}&\underset{\vec{x} \sim p_{r}(\vec{x})}{\mathbb{E}} [L(G) - \alpha \mathcal{R} (\vec{x}, H(G(F(\vec{x}))))\\ 
&+ \beta (\log I(F(\vec{x})) + \log(1 - I(G(F(\vec{x})))))]
\end{split}
\end{equation}
The result of this optimization is a purifier $G$ that not only reduces the dispersion of the confidence score vectors, but also further specializes in evading the membership classification and in maximizing the model inversion error. Besides, the utility of the target classifier is preserved.
The time overhead of predicting each input sample is incurred by the extra computation of $G$, which is a single forward pass of a neural network. Therefore, the overhead in the testing time is considered acceptable.

\section{Experimental Setup}

\subsection{Datasets}

\begin{table}[t]
	\setlength{\tabcolsep}{8pt}
	\centering
	\caption{Data allocation. A dataset is divided into training set $D_{1}$ of the target classifier, reference set $D_{2}$ and test set $D_{3}$. 
		In membership inference attack, we assume that the attacker has access to a subset $D^{A}$ of $D_{1}$ and a subset $D^{'A}$ of $D_{3}$.}
	\label{tb:datasplit}
	\begin{tabular}{c|c|r|r|c|r}
		\hline
		Dataset &  \multicolumn{1}{c|}{$D_{1}$} & \multicolumn{1}{c|}{$D_{2}$} & \multicolumn{1}{c|}{$D_{3}$} & \multicolumn{1}{c|}{$D^{A}$} & \multicolumn{1}{c}{$D^{'A}$} \\
		\hline
		CIFAR10 & 50,000 & 5,000 & 5,000 & 25,000 & 2,500\\
		Purchase100 & 20,000 & 20,000 & 20,000 & 10,000 & 10,000\\
		FaceScrub530 & 30,000 & 10,000 & 8,000 & 15,000 & 4,000\\
		\hline
	\end{tabular}
\end{table}

We use CIFAR10~\cite{shokri_membership_2017, salem_ml-leaks_2018, li_membership_2020}, Purchase100~\cite{shokri_membership_2017, nasr_machine_2018, salem_ml-leaks_2018, li_membership_2020} and 
FaceScrub530~\cite{yang_neural_2019} datasets
which are widely adopted in previous studies on model inversion and membership inference. The datasets are introduced in Appendix~\ref{app:dataset}.

Table~\ref{tb:datasplit} presents the data allocation in our experiments.
We divide each dataset into the target classifier's training set $D_1$, the reference set $D_2$ and the test set $D_3$. They have no overlap with each other. 
In membership inference attack,
we assume that the attacker has access to a subset $D^{A}$ of $D_1$ and a subset $D^{'A}$ of $D_3$.
If the attacker trains shadow models, it is assumed to have no knowledge of the membership labels of $D^{A}$ and $D^{'A}$, and use half of them to train the shadow models. 
The attack model is trained on the whole $D^A$ and $D^{'A}$ with the shadow models' training data labeled as members.
Otherwise, the attacker is assumed to have the membership labels and can directly query the target classifier to get the confidence scores for members and non-members.
We use the remaining data in $D_1$ and $D_3$ to test the membership inference accuracy.
We use $D_2$ as the reference set for defenses that require it in their mechanisms, e.g., MemGuard~\cite{jia_memguard_2019}, Min-Max~\cite{nasr_machine_2018} and our approach.
In the model inversion attack, for the FaceScrub530 classifier, the attacker uses a CelebA dataset as the auxiliary dataset to train the inversion model, which is the same setting in \cite{yang_neural_2019}. For other classifiers, the attacker samples 80\% from $D_1$, $D_2$ and $D_3$ respectively to compose the auxiliary dataset, and uses the remaining 20\% data to test the inversion error.

\subsection{Target Classifier}

For the CIFAR10 dataset, we use the DenseNet121 architecture~\cite{huang_densely_2017}, which is also used in \cite{nasr_machine_2018, li_membership_2020}. 
We train our classifier with stochastic gradient descent (SGD) optimizer for 350 epochs with learning rate 0.1 from epoch 0 to 150, 0.01 from 150 to 250, and 0.001 from 250 to 350. The classifier is regularized with 
$L_2$ regularizaiton (weight decay parameter 5e-4).
For the Purchase100 dataset, we use the same model and training strategy as in \cite{nasr_machine_2018} to train the target classifier. It is a 4-layer fully connected neural network.
For the FaceScrub530 dataset, we use the same convolutional neural network and the same training strategy as in \cite{yang_neural_2019} to train the target classifier.

\subsection{Purification Framework}

The purification framework consists of the purifier, the adversarial model and the discriminator.
We introduce their model architectures in Appendix~\ref{app:purification}.

\subsection{Data Inference Attacks}

In our experiments, we consider the following model inversion and membership inference attacks as introduced in Section~\ref{sec:datainference}.

\noindent \textbf{Model inversion attack~\cite{yang_neural_2019}.} 
We adopt the black-box model inversion attack~\cite{yang_neural_2019}, where the attacker trains an inversion model to infer the reconstruction of the input sample. We use the same model architecture as in \cite{yang_neural_2019} to train the inversion model for FaceScrub530 dataset. The inversion model was trained on the CelebA dataset which is the same auxiliary dataset used in \cite{yang_neural_2019}.

\noindent \textbf{Mlleaks attack~\cite{salem_ml-leaks_2018}.} 
This is a confidence-based membership inference attack, where the attacker has no knowledge of the membership labels of $D^A$ and $D^{'A}$. Therefore, it has to train a shadow model to replicate the target classifier and then trains the attack model based on the confidence scores of the shadow model.
To consider the strongest attack, we assume that the shadow model has the same architecture as the target classifier.
We use a multi-layer perceptron with a 128-unit hidden layer and a sigmoid output layer to train the attack model. 
We use the Adam optimizer with learning rate 0.001. The number of training epochs is set to 50 for each dataset.

\noindent \textbf{Mlleaks-a attack~\cite{salem_ml-leaks_2018}.} This is an adaptive version of the Mlleaks attack, where the attacker is assumed to know the defender's purification framework. Hence, he can train his own purification framework on $D_2$ to increase the attack accuracy. 

\noindent \textbf{NSH attack~\cite{nasr_machine_2018}.} This is a confidence \& label-based membership inference attack proposed by Nasr, Shokri and Houmansadr~\cite{nasr_machine_2018}. The attacker is assumed to have the knowledge of the membership labels of $D^A$ and $D^{'A}$, and thus can directly query the target classifier to get the confidence score vectors of members and non-members without training the shadow model.
We use the same attack model as in \cite{nasr_machine_2018} for our implementation.
During the training of the attack model,
we make sure every training batch has the same number of member and non-member instances to prevent the attack model to be biased toward one side as \cite{nasr_machine_2018} did.

\noindent \textbf{Label attack~\cite{yeom_privacy_2018}.} This is a label-based membership inference attack, where the attacker predicts an input sample is a member if and only if it is correctly classified by the target classifier. 

\subsection{Existing Defenses}
\label{sec:existing}

We compare our approach with the following defense methods most of which are introduced in Section~\ref{sec:existing_defense}.

\noindent \textbf{Min-Max~\cite{nasr_machine_2018}.}
We use the open-source code of \cite{nasr_machine_2018} to implement the Min-Max defense. The number of training epochs for both the classification model and the inference model is the same as that we use to train the target classifier. 

\noindent \textbf{MemGuard~\cite{jia_memguard_2019}.}
We adopt the open-source code of \cite{jia_memguard_2019} to implement this defense method. Specifically, the defense classifier used in this method consists of three hidden layers [256, 128, 64] and uses ReLU in hidden layers and Sigmoid in the output layer.

\noindent \textbf{Model-Stacking~\cite{salem_ml-leaks_2018}.}
This method ensembles two layers of models. The first layer combines two models, where we use the same architecture as the target classifier as the first model and use random forest for Purchase100 and VGG19~\cite{simonyan2014very} for FaceScrub530 and CIFAR10 as the second model. 
The second layer is a logistic regression model for Purchase100 and FaceScrub530, and a neural network with a single hidden layer of size 128 for CIFAR10.

\noindent \textbf{One-Hot Encoding.} It encodes a confidence score vector into a one-hot vector (i.e., the entry with the largest confidence is set to 1 and the other entries are all 0). 

\noindent \textbf{Random Noise.} We follow \cite{jia_memguard_2019} to add random noise to the confidence score vectors as a defense method. It randomly modifies all the confidence scores but is constrained to predict the correct label.

\subsection{Metrics}

We use the following metrics to measure the utility, defense performance and efficiency of a defense method.

\noindent \textbf{Classification Accuracy.} It is measured on the training set $D_1$ and the test set $D_3$ of the target classifier. It reflects how good the target classifier is on the classification task. 

\noindent \textbf{Confidence Score Distortion.} We measure the confidence score distortion introduced by a defense method by computing the $L_2$ norm of the distance between the original confidence score vector predicted by the target classifier and the new confidence score vector after the defense method is applied.
We report the relative distortion divided by the sum (i.e., 1.0) of the elements in the confidence score vector.

\noindent \textbf{Inference Accuracy.} This is the classification accuracy of the attacker's attack model in predicting the membership of input samples. It is measured on $D_1 - D^A$ (i.e., members) and $D_3 - D^{'A}$ (i.e., non-members).

\noindent \textbf{Inversion Error.} We measure the inversion error by computing the mean squared error ($L_2$ norm) between the original input sample and the reconstructed sample. For the FaceScrub530 classifier, it is measured on $D_1$ and $D_3$. For other classifiers, it is measured on the 20\% of $D_1$ and $D_3$ respectively.

\noindent \textbf{Efficiency.} We measure the efficiency of a defense method by reporting the relative training time and testing time versus the original target classifier.

\section{Experimental Results}

\subsection{Defense Performance of Purifier}

\begin{figure*}[t]
	\centering
	\begin{minipage}[b]{1\linewidth}
		\centering
		\subfigure{
			\includegraphics[width=1\linewidth]{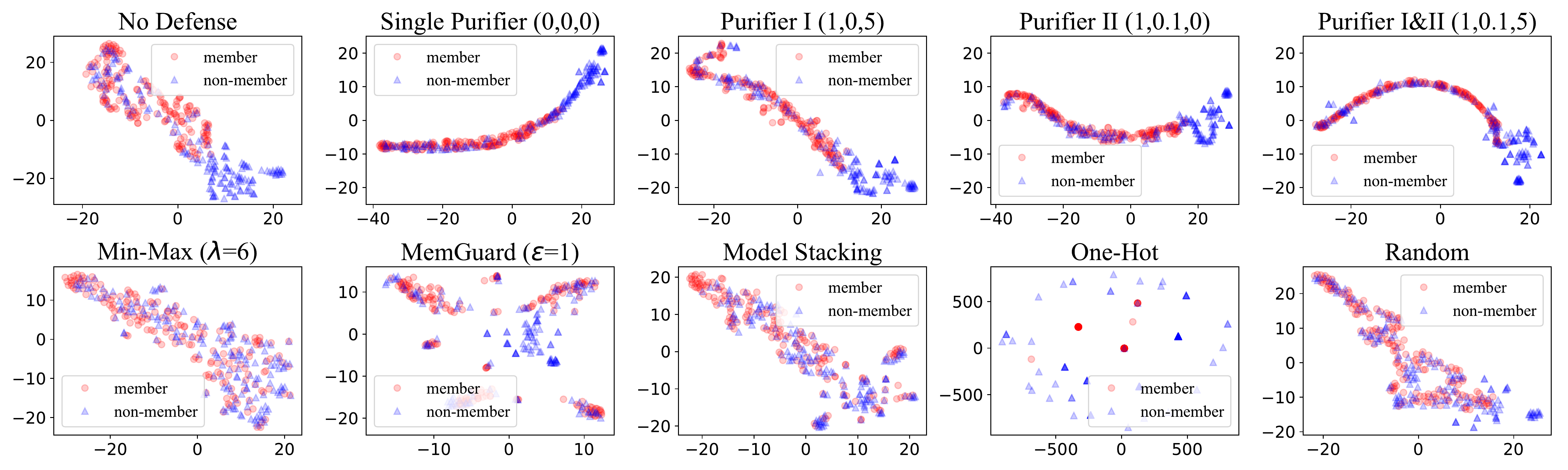}
		}
	\end{minipage}
	\caption{The confidence score vectors on members and non-members projected into a 2D space using t-Distributed Stochastic Neighbor Embedding (t-SNE) for each defense. The confidence score vectors are predicted on data from the class 74 of the Purchase100 dataset. The purifier makes confidence score vectors have very clear patterns, and also reduces their dispersion.}
	\label{fig:tsne}
\end{figure*}

\begin{table*}[t]
	\setlength{\tabcolsep}{6pt}
	\centering
	\caption{Defense performance of different purification architectures. }
	\label{tb:bigtable}
	\begin{tabular}{l|l|r|r|c|c|r|r|c|r}
		\hline
		\multicolumn{1}{c|}{Dataset} & \multicolumn{1}{c|}{Defense} & \multicolumn{3}{c|}{Utility} & \multicolumn{1}{c|}{Inversion Error} & \multicolumn{4}{c}{Inference Accuracy}\\
		\hhline{~|~|-|-|-|-|-|-|-|-}
		&  & Train acc. & Test acc. & Conf. dist. & $L_2$ norm & NSH & Mlleaks & Mlleaks-a & Label \\
		\hline
		\multirow{5}{*}{CIFAR10} & None & 99.99\% & 95.92\% & 0 & 1.4357 & 56.03\% & 56.26\% & -- & 52.20\% \\
		& Single Purifier (0,0,0) & 100.00\% & 95.78\% & 0.66\% & 1.4785 & 53.85\% &51.37\% &52.06\% &52.28\%\\
		& Purifier \rom{1} (1,0,10) & 100.00\% & 95.80\% & 0.87\% & 1.4941 & 53.21\% & 51.52\% & 51.28\% &52.28\%\\
		& Purifier \rom{2} (1,0.1,0) & 100.00\% & 95.82\% & 0.66\% & 1.4785 & 52.89\% & 51.78\% & 51.21\% &52.30\%\\
		& Purifier \rom{1} \& \rom{2} (1,1,0.1) & 100.00\% & 95.76\% & 5.42\% & 1.4940 & 51.25\% & 51.30\% & 50.82\% &52.32\%\\
		\hline 
		\multirow{5}{*}{Purchase100} & None & 100.00\% & 84.36\% & 0 & 0.1426 & 70.36\% & 64.43\% & -- & 57.96\%\\
		& Single Purifier (0,0,0) & 99.96\% & 84.21\% & 1.86\% & 0.1518 & 67.40\% & 57.92\% & 59.40\% & 57.88\% \\
		& Purifier \rom{1} (1,0,5) & 99.96\% & 84.17\% & 3.52\% & 0.1521 & 58.93\% & 52.45\% & 56.83\% & 57.90\%\\
		& Purifier \rom{2} (1,0.1,0) & 100.00\% & 84.23\% & 3.81\% & 0.1518 & 58.40\% & 52.06\%& 55.50\% & 57.89\%\\
		& Purifier \rom{1} \& \rom{2} (1,0.1,5) & 99.96\% & 84.17\% & 4.38\% & 0.1521 & 58.13\% & 51.31\% & 52.82\%& 57.90\%\\
		\hline
		\multirow{5}{*}{FaceScrub530} & None & 100.00\% & 77.68\% & 0 & 0.0114 & 69.34\% & 75.04\% & -- & 61.36\%\\
		& Single Purifier (0,0,0) & 99.54\% & 77.30\% & 2.49\% & 0.0443 & 65.20\% & 64.20\% & 65.14\% & 61.12\% \\
		& Purifier \rom{1} (0.1,0,1) & 99.88\% & 77.48\% & 3.01\% & 0.0448 & 65.14\% & 63.20\% & 61.87\% & 61.45\%\\
		& Purifier \rom{2} (1,1,0) & 99.97\% & 77.40\% & 4.07\% & 0.0446 & 57.99\% & 58.13\%& 61.41\% & 61.29\%\\
		& Purifier \rom{1} \& \rom{2} (0.1,0.1,5) & 99.97\% & 77.49\% & 4.38\% & 0.0448 & 58.14\% & 60.40\% & 60.68\%& 61.34\%\\
		\hline
	\end{tabular}
\end{table*}

\begin{figure}[t]
	\centering   
	\subfigure{
		\includegraphics[width=0.93\linewidth]{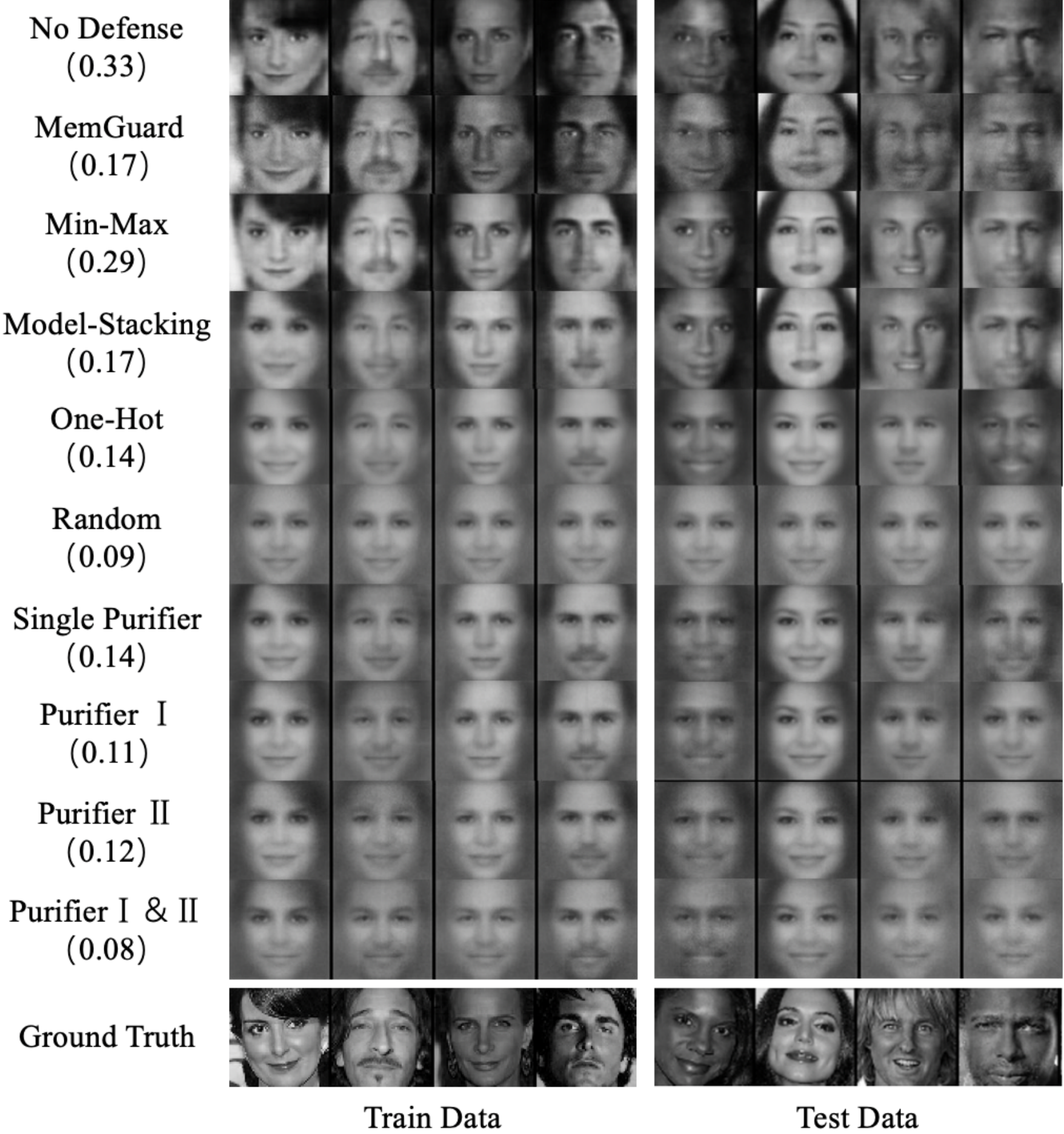}
	}
	\caption{Model inversion attack against the FaceScrub530 classifier defended by different approaches.}
	\label{fig:facescrub}
\end{figure}

\noindent \underline{\textbf{Purifier is Effective in Mitigating Data Inference}}

To have a better knowledge of how each component of the purification framework performs in mitigating data inference attacks, we present the defense performance when the single purifier is used, the purifier and the adversarial model are used (Purifier \rom{1}), the purifier and the discriminator are used (Purifier \rom{2}), and all the three models are used (Purifier \rom{1} \& \rom{2}) in Table~\ref{tb:bigtable}. 
The 3-tuple numbers appended to each purifier represent the hyper-parameter $\lambda$, $\alpha$ and $\beta$ respectively.
We train each purification architecture for three times and report the average experimental results in Table~\ref{tb:bigtable}.
For each classification task, we can see that the single purifier is able to concurrently decrease the inference accuracy and increase the inversion error as well as to preserve the classification accuracy. 
To gain a better insight of how the single purifier works, we visualize the confidence score vectors by projecting them in the 2D space using t-Distributed Stochastic Neighbor Embedding (t-SNE)~\cite{tsne} in Figure~\ref{fig:tsne}. 
We can see that, without defense, the confidence score vectors are widely scattered. After purification, they are more tightly clustered and have clear patterns no matter for members or non-members.
This demonstrates that the purifier is able to reduce the dispersion of the confidence score vectors, which is helpful to defend both the membership inference and the model inversion attacks.

When the purifier is trained with the adversarial model in Purifier \rom{1}, i.e., the purifier is specialized in defending the model inversion attack,  the inversion error further increases as expected. It is interesting to note that, the inference accuracy also decreases even though the purifier is dedicated to model inversion. A similar conclusion can be reached when the purifier is trained with the discriminator in Purifier \rom{2}. The inference accuracy further decreases while the inversion error could also increase, for example, on the FaceScrub530 dataset.
This finding demonstrates again that the two attacks are actually related. 
Defending one attack can naturally lead to the defense against the other one. 
When the purifier is jointly trained with both the adversarial model and the discriminator in Purifier \rom{1} \& \rom{2}, it achieves the best overall performance against both attacks. 
For example, the inversion error is the largest and the inference accuracy is the lowest on the CIFAR10 and Purchase530 datasets.
We visualize the confidence score vectors purified by different architectures in Figure~\ref{fig:tsne} (row 1).
The results show that no matter which purification framework is used, the dispersion of confidence score vectors is consistently reduced. We believe it is the common reason of how the purification framework can defend the two attacks at the same time.
Besides defense, the purification incurs negligible utility loss. For instance, it imposes less than 0.4\% classification accuracy drop and less than 5.5\% confidence score distortion for all the evaluated tasks.

To gain an intuitive understanding of the defense performance against the model inversion attack, we present the reconstructed facial images by the attacker on the FaceScrub530 dataset in Figure~\ref{fig:facescrub} (row 7-10). 
The ground truth images are single images that are randomly selected from the the training data $D_1$ (members) and the test data $D_3$ (non-members) of the FaceScrub530 classifier. The attacker aims to reconstruct them from the confidence scores in a way described in~\cite{yang_neural_2019}.
The numbers appended to the defense name are the average facial similarity scores compared with the ground truth, which we obtained from the Microsoft Azure Face Recognition service\footnote{https://azure.microsoft.com/en-us/services/cognitive-services/face/}. 
Without defense, the attacker can infer a very accurate reconstruction of the facial images, which demonstrates that the model inversion attack works well. However, when the purifier is applied, the reconstructed facial images lose much information of the ground truth image, which can be also demonstrated by the much lower similarity score.

It is worth noting that the purifier has almost no defense effect against the Label attack where the confidence score information is not leveraged. Essentially, the inference accuracy of this attack can be estimated as $0.5+g/2$ where $g$ is the generalization error, i.e., difference between the training accuracy and the testing accuracy of the target classifier~\cite{li_membership_2020}. 
The purifier does not change the generalization error too much but, on the other hand, examines to which extent the confidence score information can be purified such that the attacker cannot exploit them to gain additional useful information. For example, the NSH attack exploits both label and confidence information to infer the membership. The purifier is able to reduce its inference accuracy to the comparable level to the Label attack, which demonstrates that the attacker cannot leverage the purified confidence scores to gain more membership information than what the label already leaks. 
Note that the NSH accuracy against the FaceScrub530 classifier defended by the purifier is lower than the Label attack. This is due to the limited amount of data that the attack has to train the NSH classifier. When we increase the attacker's data amount to 80\% of the target classifier's training data, the NSH accuracy can achieve 63\%. However, having so much member data might not be practical for the attacker.
Table~\ref{tb:label_attack} presents the performance of existing methods against the Label attack. 
We can see that the purifier has comparable defense performance with existing methods.
The purifier can be applied together with regularization techniques to reduce the generalization error to mitigate Label attack, which we leave as future work.

\begin{table}[t]
	\centering
	\setlength{\tabcolsep}{1pt}
	\caption{Inference accuracy of Label attack against different defense methods.}
	\label{tb:label_attack}
	\begin{tabular}{c|c|c|c|c|c|c|c}
		\hline
		Dataset & None & Purifier & MinMax & MemGuard & ModelStacking & OneHot & Random   \\
		\hline
		CIFAR10 & 52.20\% & 52.32\% & 52.65\% & 52.20\% & 51.86\% & 52.20\% & 52.20\% \\
		\hline
		Purchase100 & 57.96\% & 57.90\% & 56.12\% & 57.96\% & 56.01\% & 57.96\% & 57.96\% \\
		\hline
		FaceScrub530 & 61.36\% & 61.34\% & 66.01\% & 61.36\% & 64.93\% & 61.36\% & 61.36\% \\
		\hline
	\end{tabular}
\end{table}

\noindent \underline{\textbf{Closer Look at Confidence Scores}}

\begin{figure}[t]
	\centering
	\begin{minipage}[b]{1\linewidth}
		\centering
		\subfigure[Train data]{
			\label{fig:mem_soft_label}
			\includegraphics[width=0.5\linewidth]{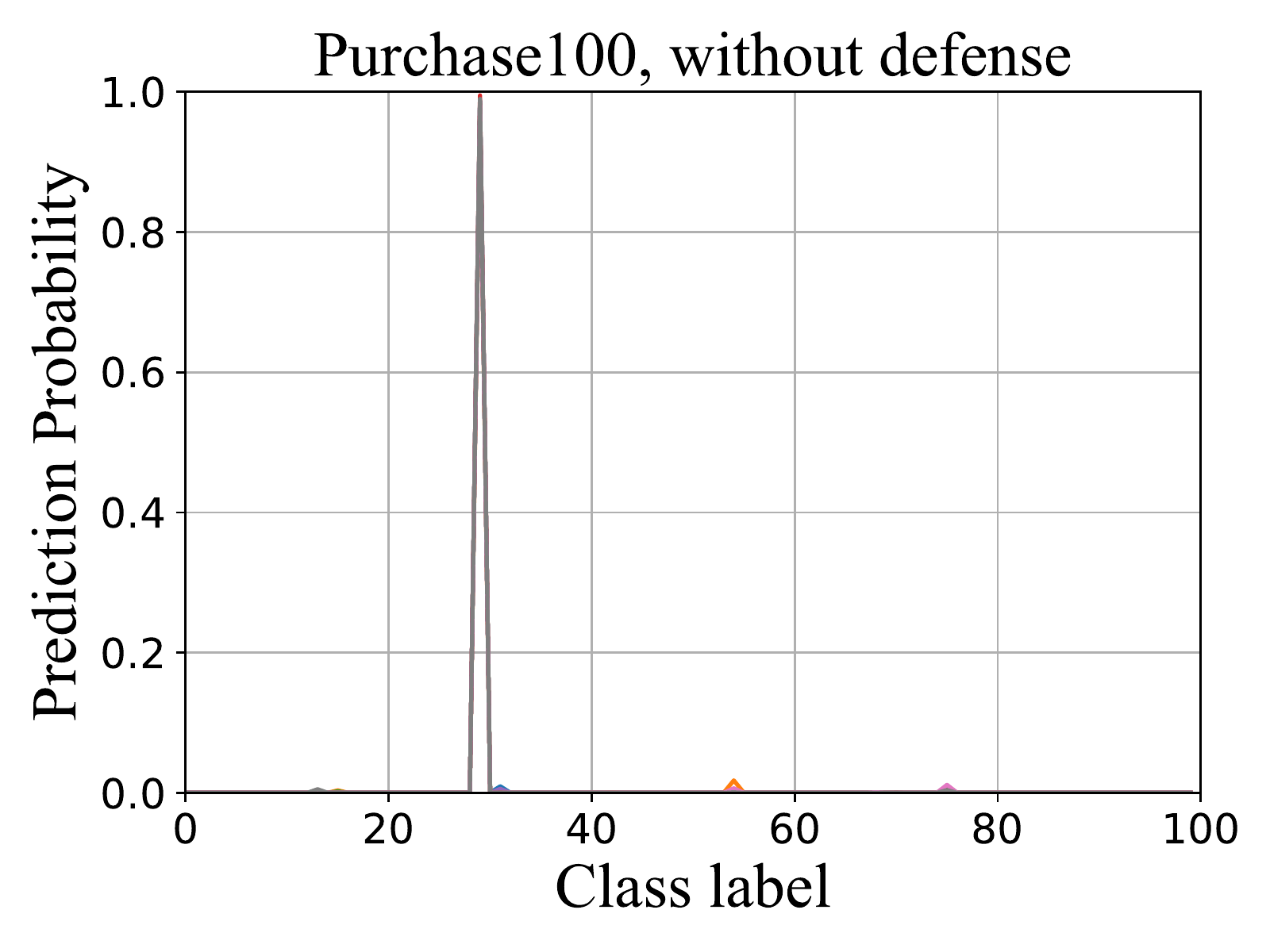}
			\includegraphics[width=0.5\linewidth]{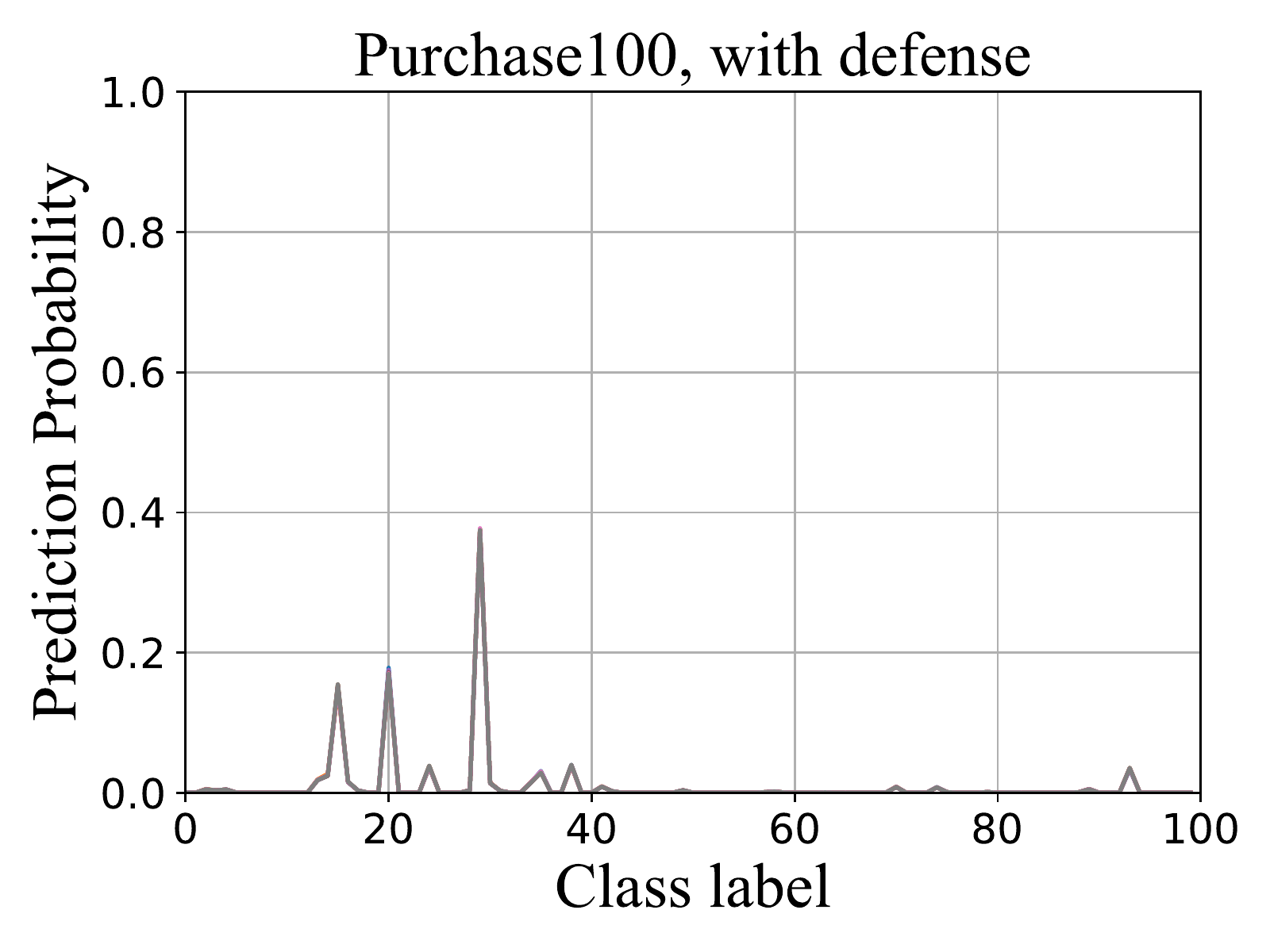}
		}
		
		\subfigure[Test data]{
			\label{fig:nonmem_soft_label}
			\includegraphics[width=0.5\linewidth]{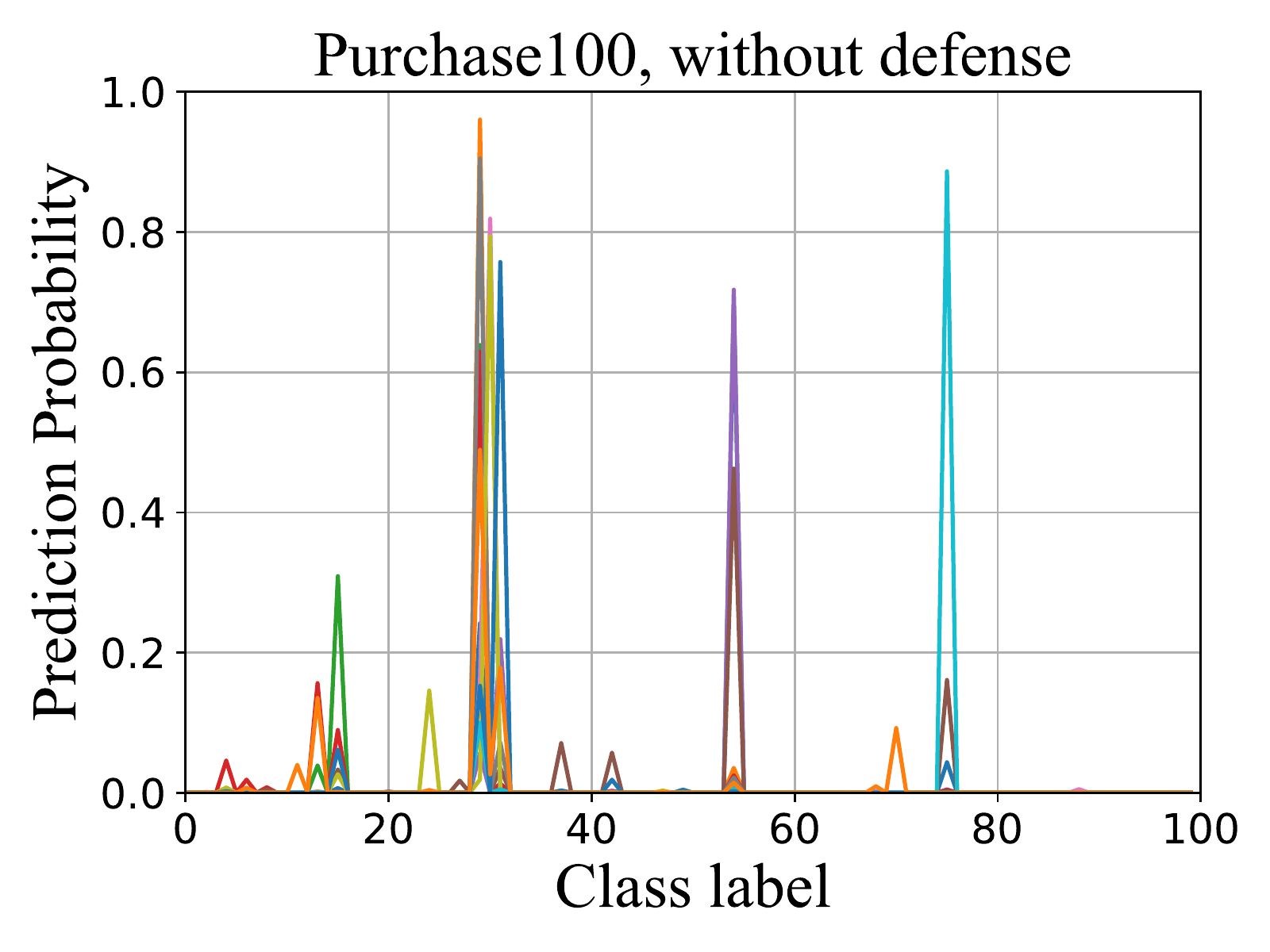}
			\includegraphics[width=0.5\linewidth]{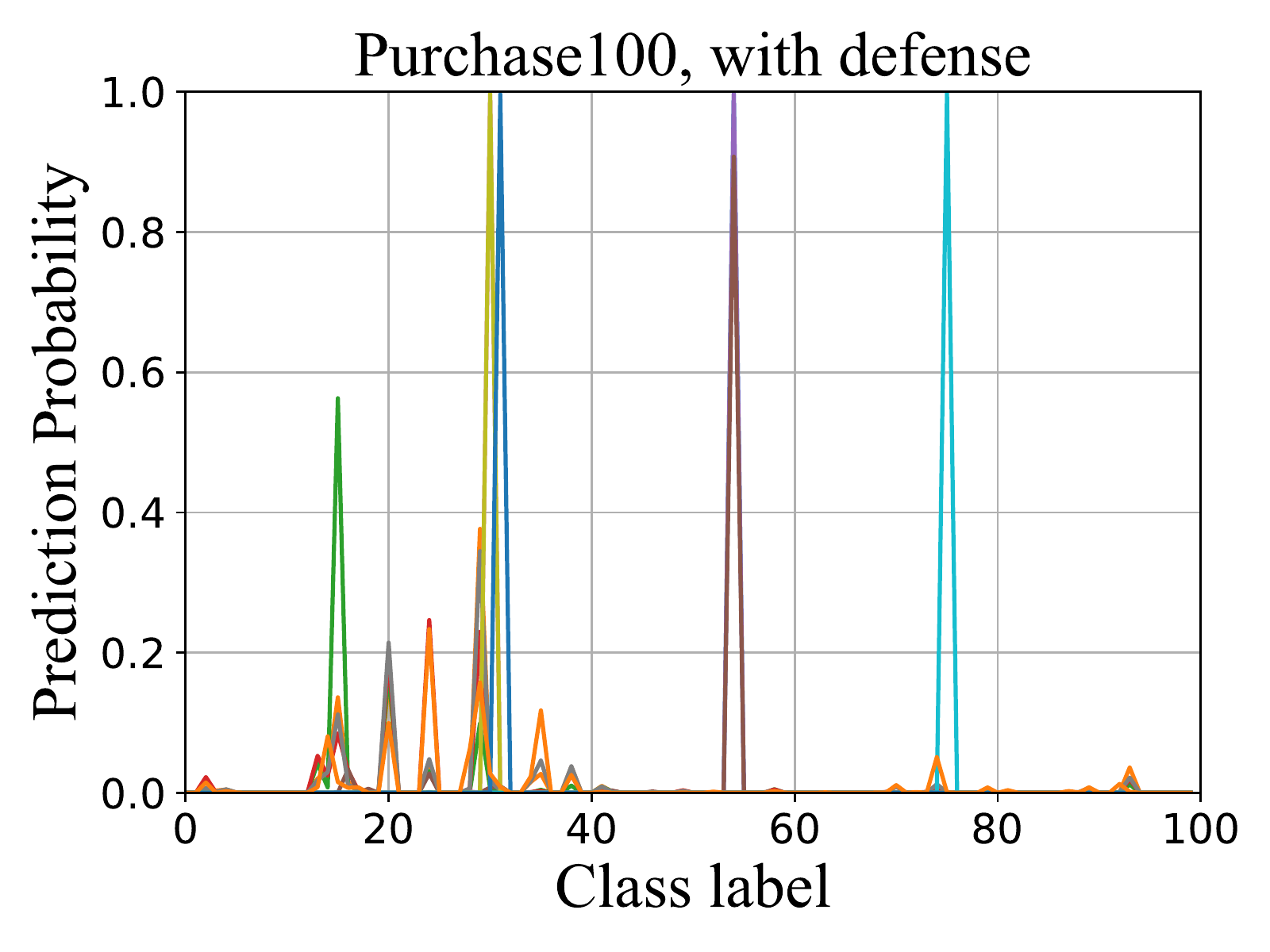}
		}
	\end{minipage}
	\caption{Distribution of the confidence score vectors of the target classifier on the training data and test data of class 29 in the Purchase100 dataset. Each color represents one data record.}
	\label{fig:soft_label}
\end{figure}

\begin{figure}[t]
	\centering
	\begin{minipage}[b]{1\linewidth}
		\centering
		\subfigure{
			\includegraphics[width=0.5\linewidth]{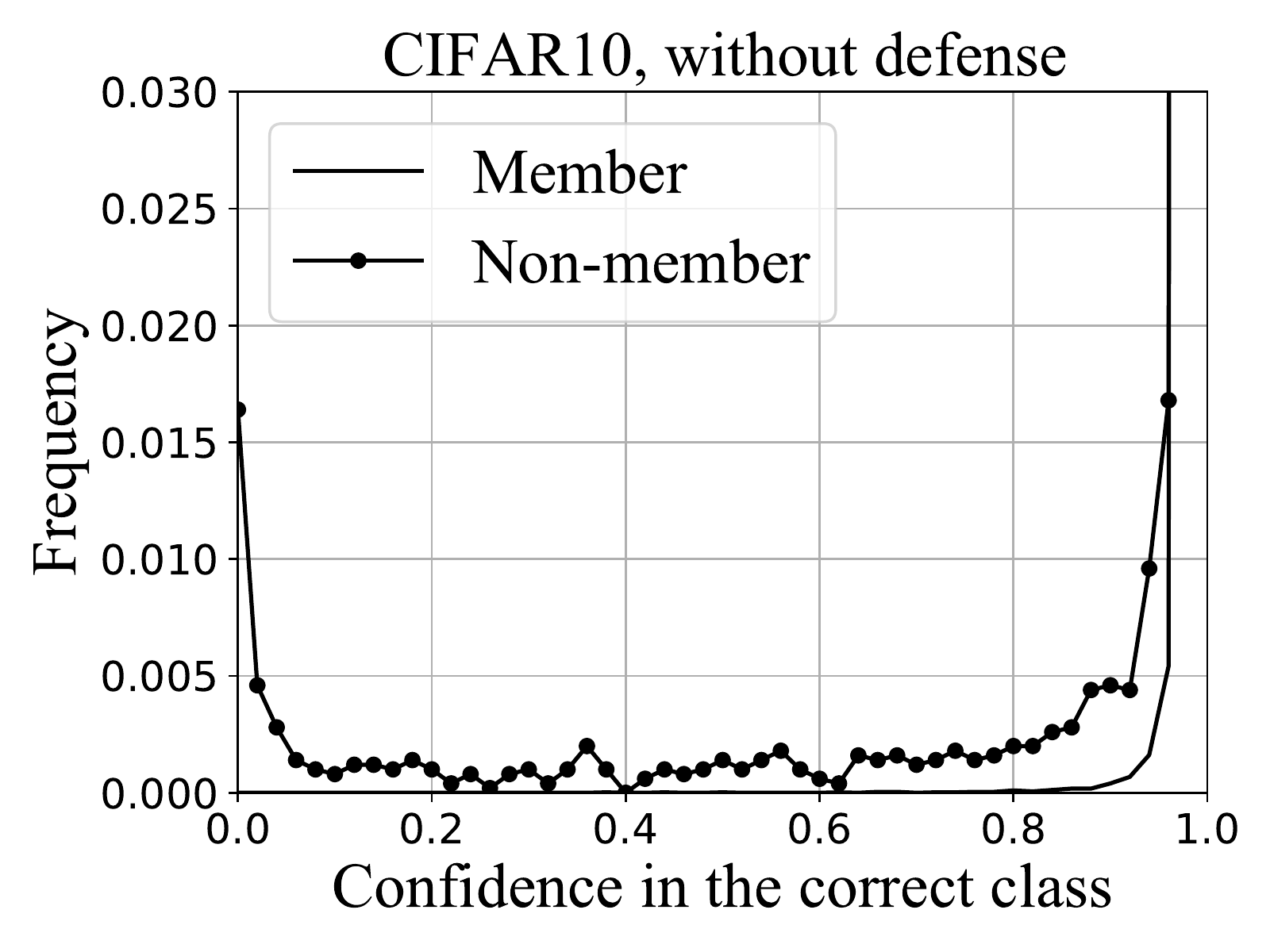}
			\includegraphics[width=0.5\linewidth]{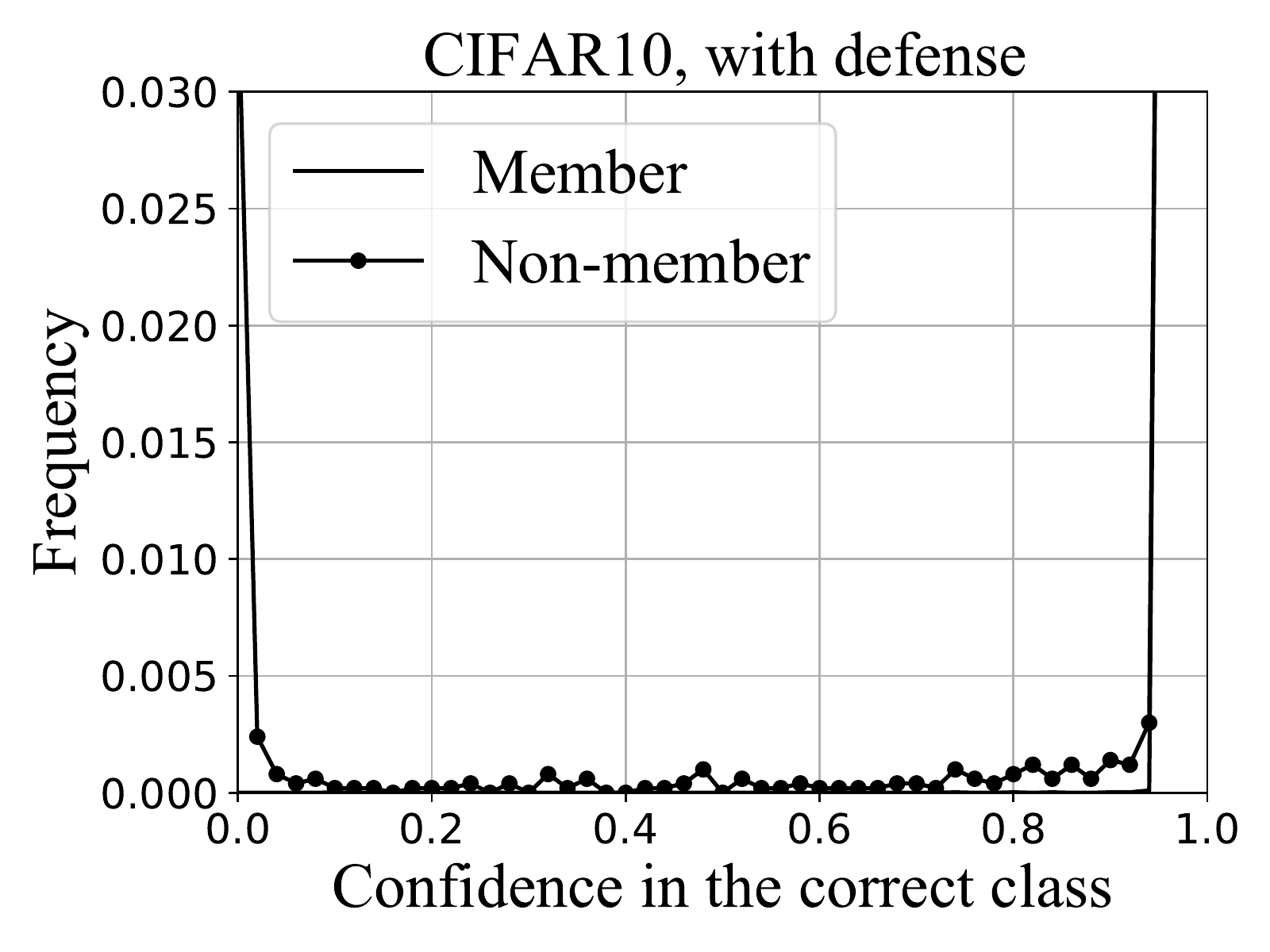}
		}
		
		\subfigure{
			\includegraphics[width=0.5\linewidth]{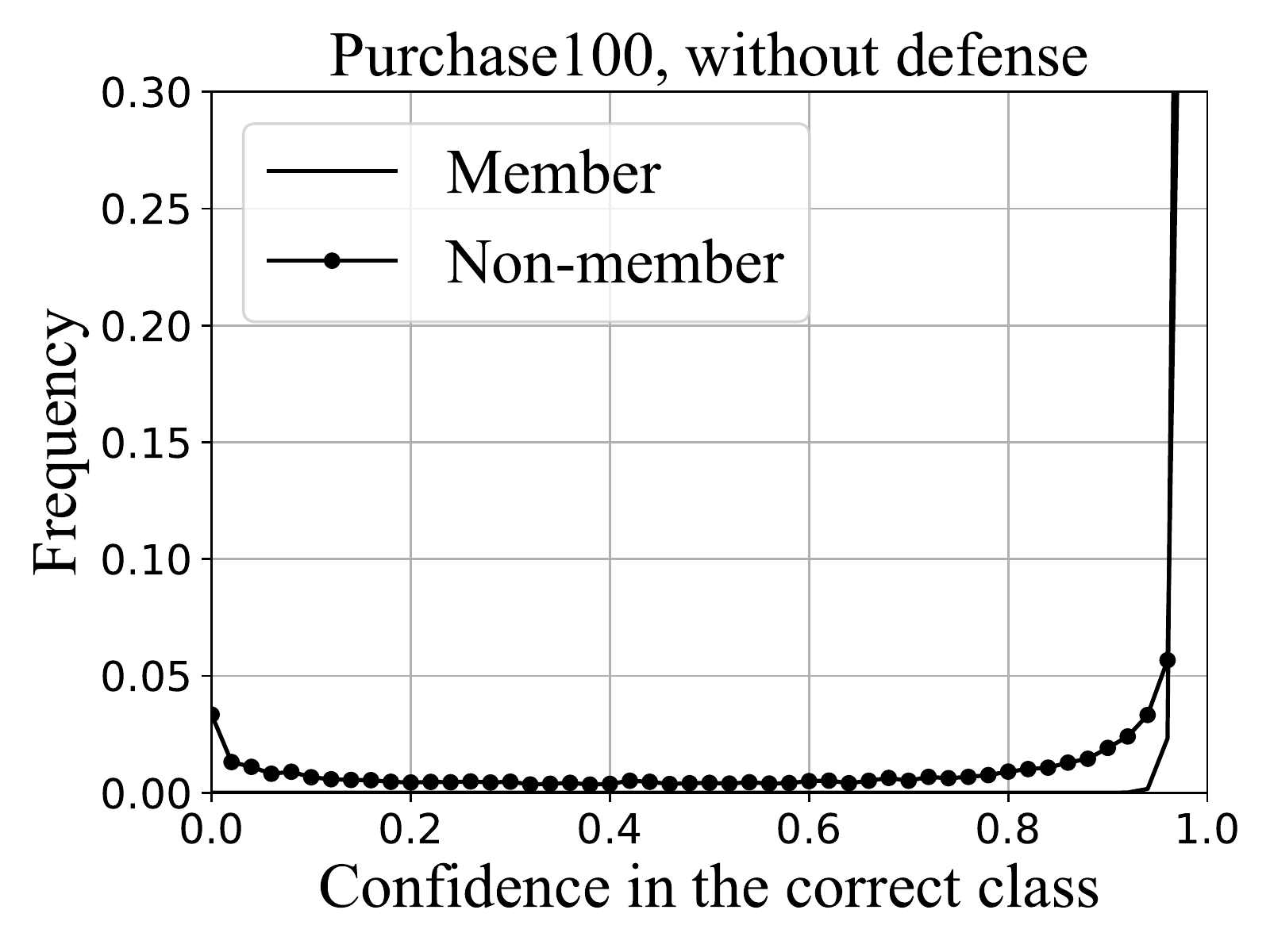}
			\includegraphics[width=0.5\linewidth]{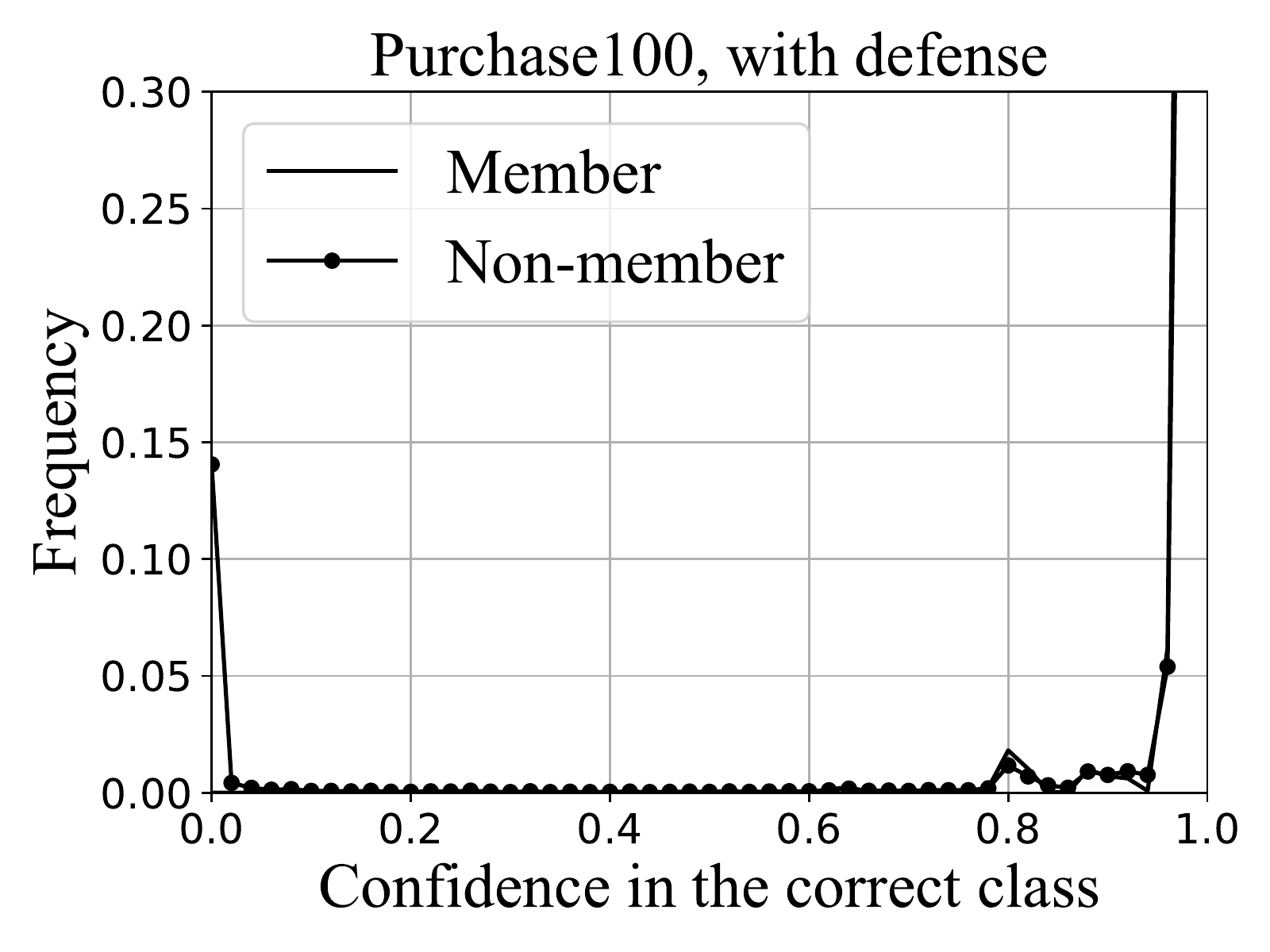}
		}
		
		\subfigure{
			\includegraphics[width=0.5\linewidth]{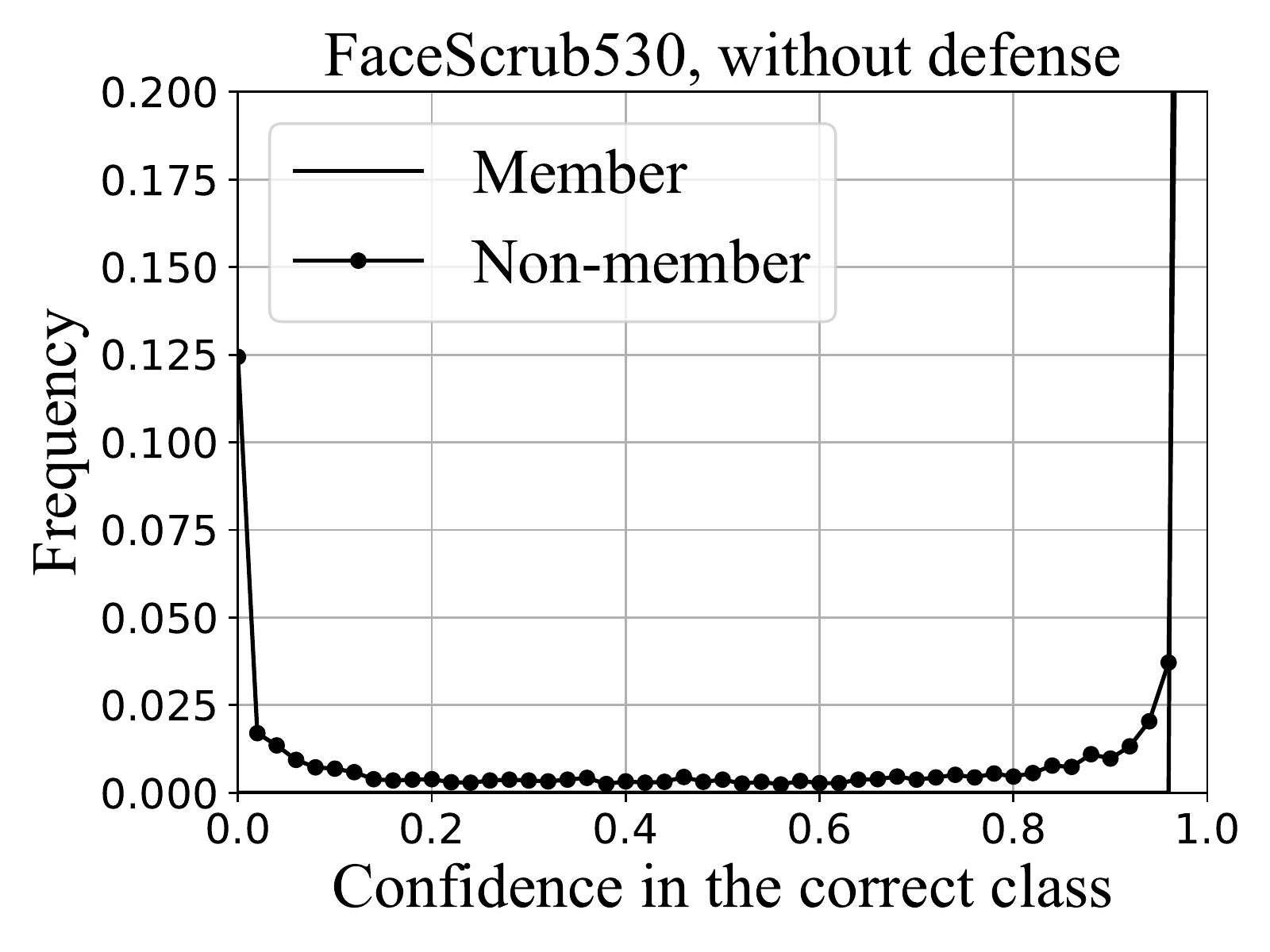}
			\includegraphics[width=0.5\linewidth]{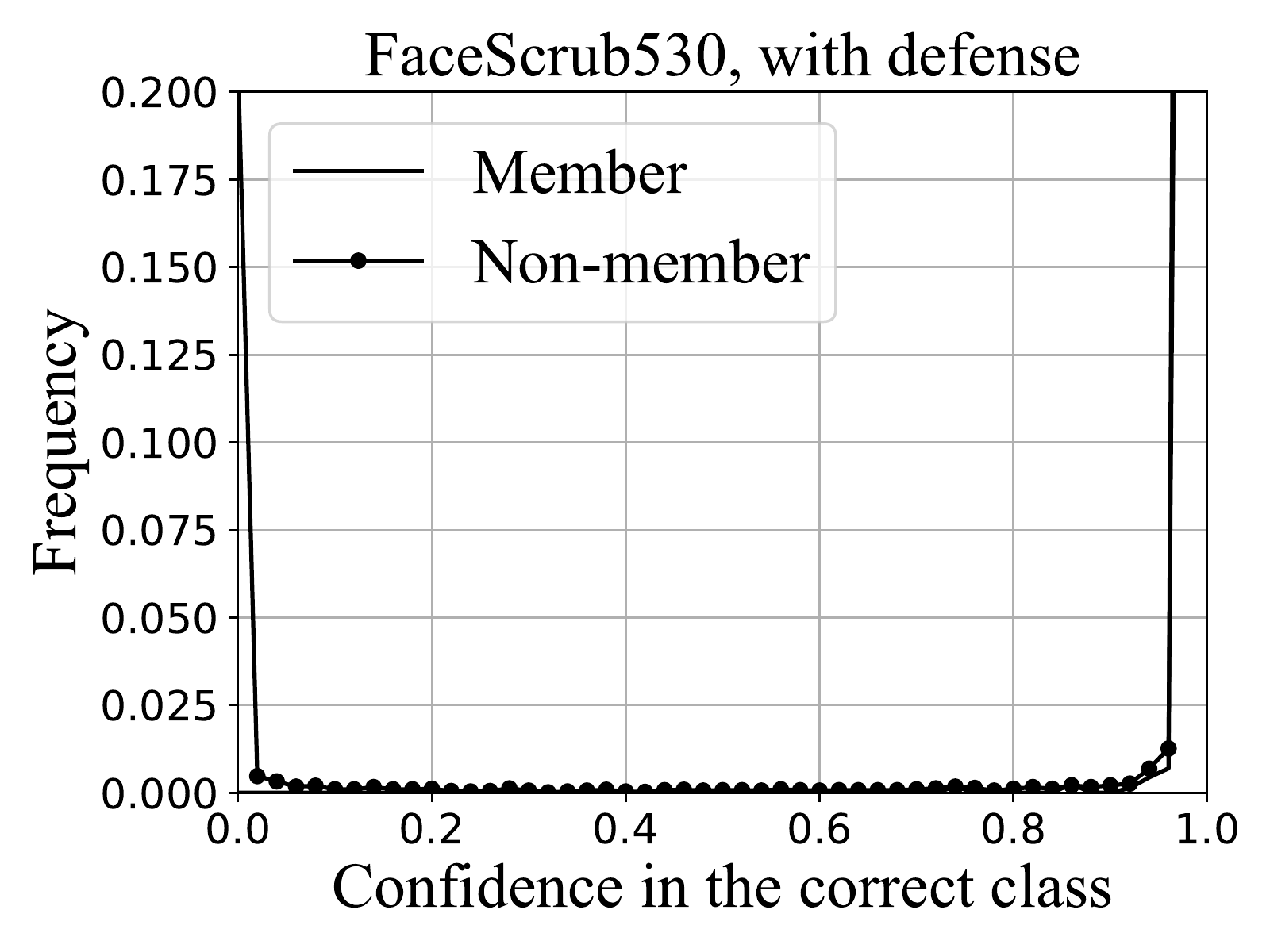}
		}

	\end{minipage}
	
	\caption{Distribution of the target classifier's confidence in predicting the correct class on members and non-members of its training set.}
	\label{fig:prediction_acc}
\end{figure}

\begin{figure}[t]
	\centering
	\begin{minipage}[b]{1\linewidth}
		\centering
		\subfigure{
			\includegraphics[width=0.5\linewidth]{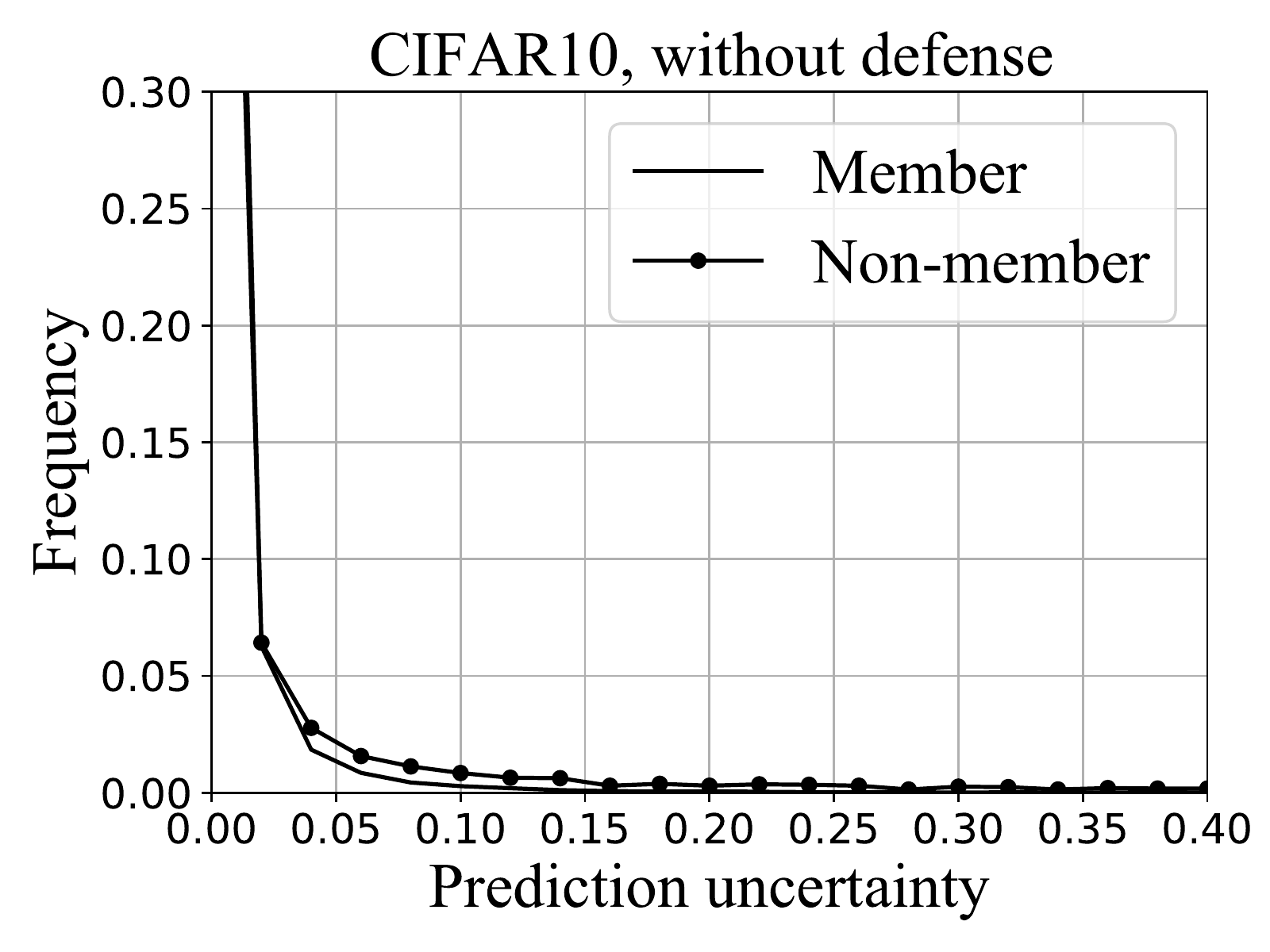}
			\includegraphics[width=0.5\linewidth]{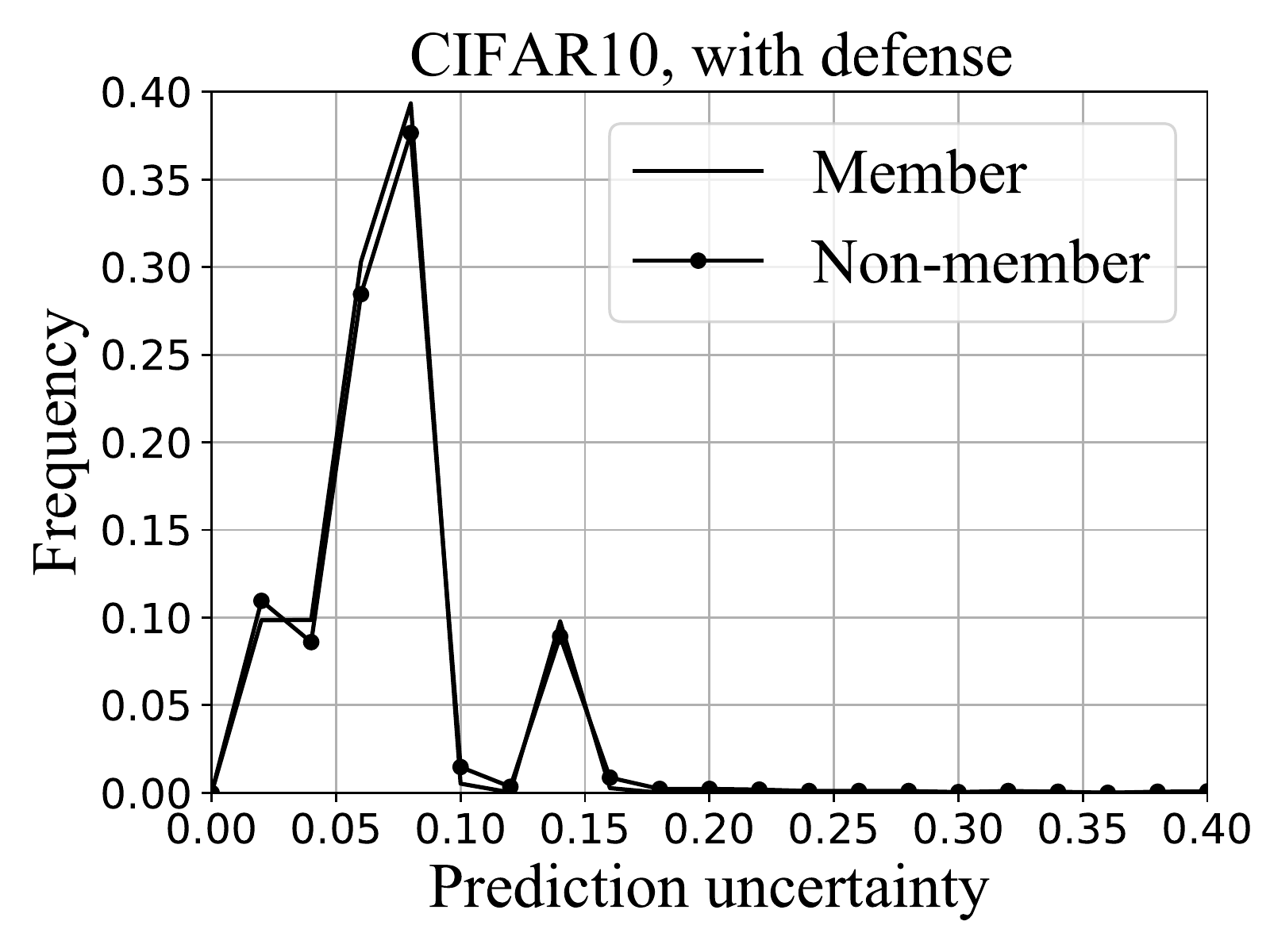}
		}
		
		\subfigure{
			\includegraphics[width=0.5\linewidth]{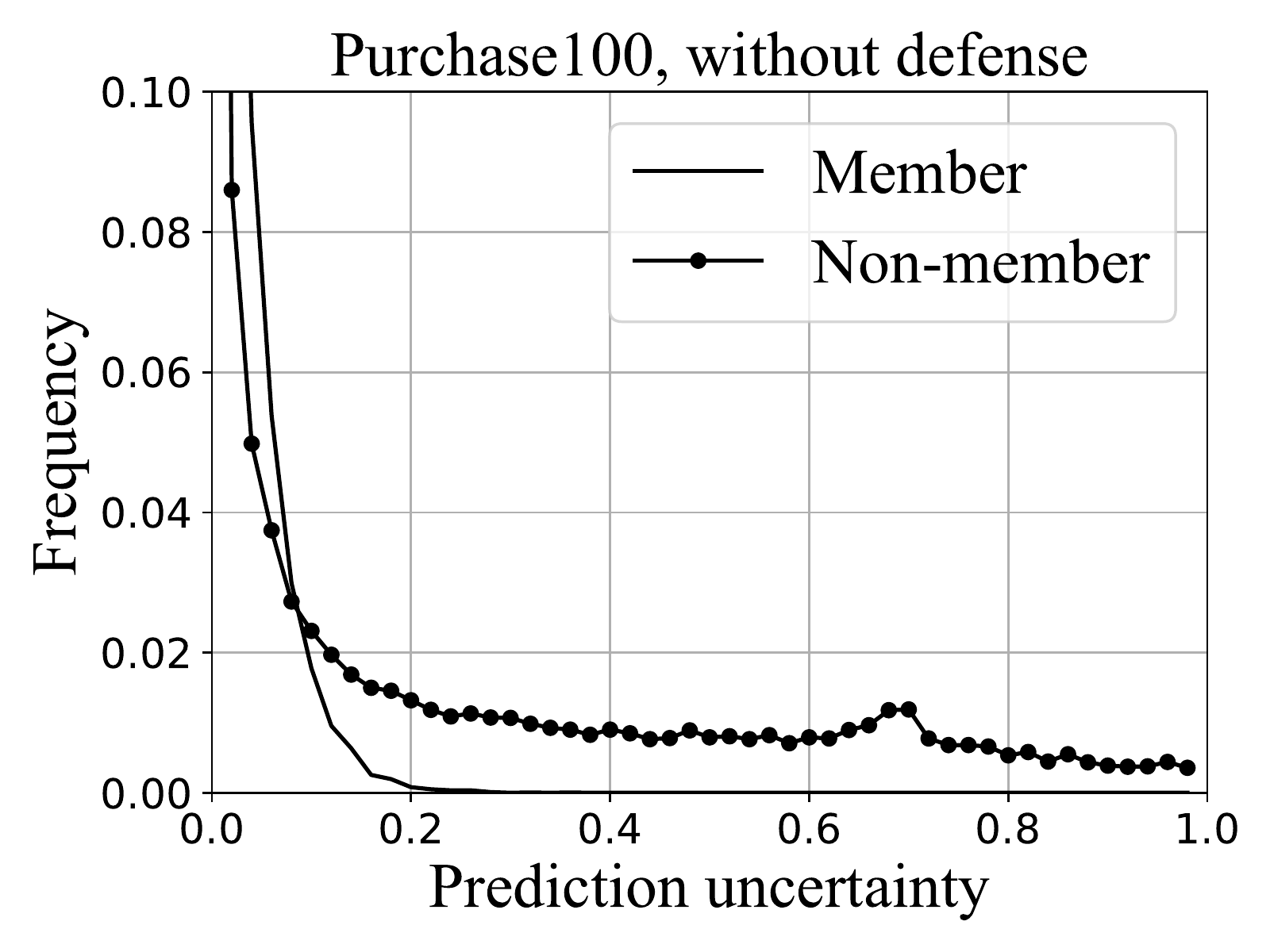}
			\includegraphics[width=0.5\linewidth]{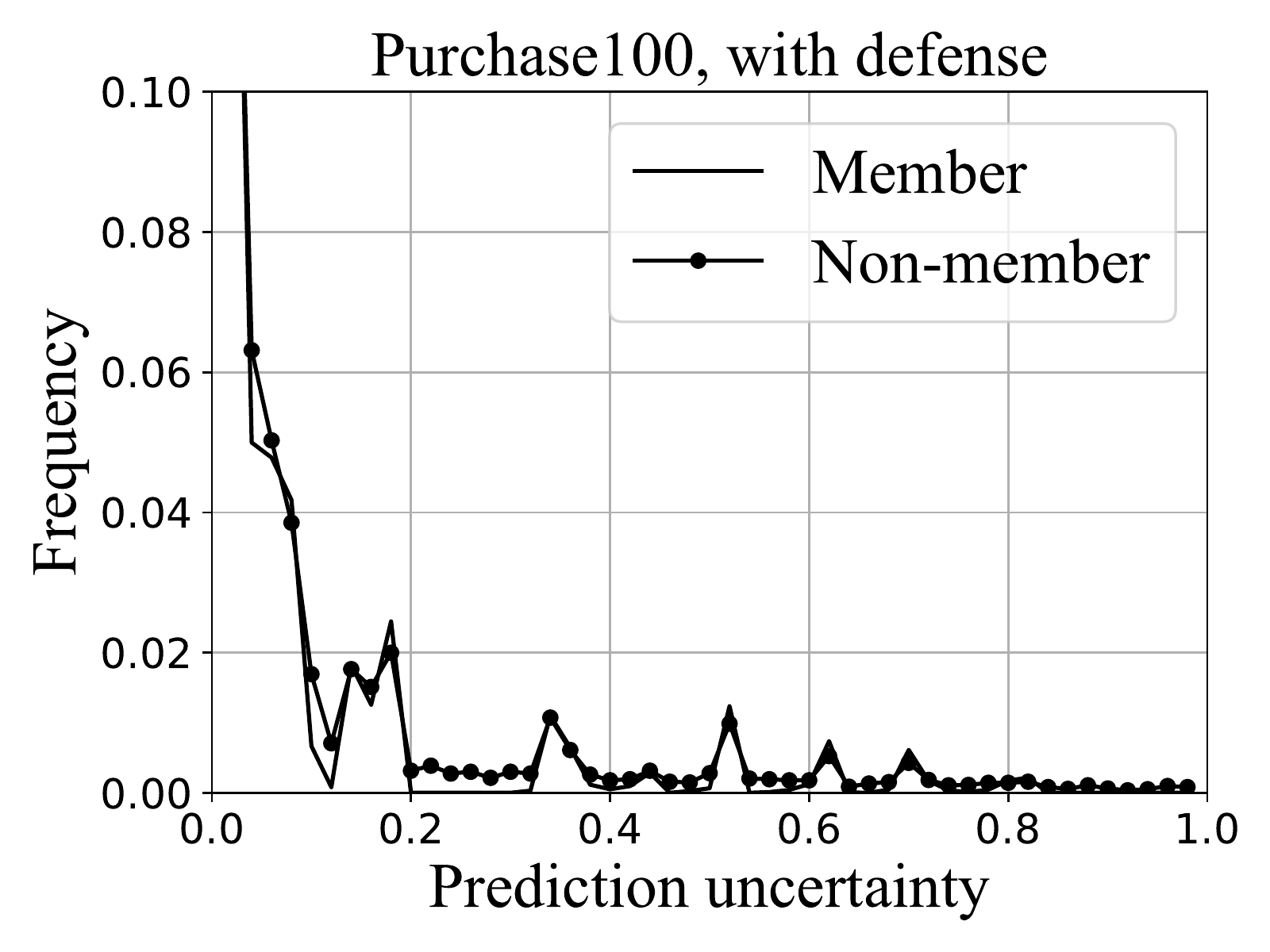}
		}
		
		\subfigure{
			\includegraphics[width=0.5\linewidth]{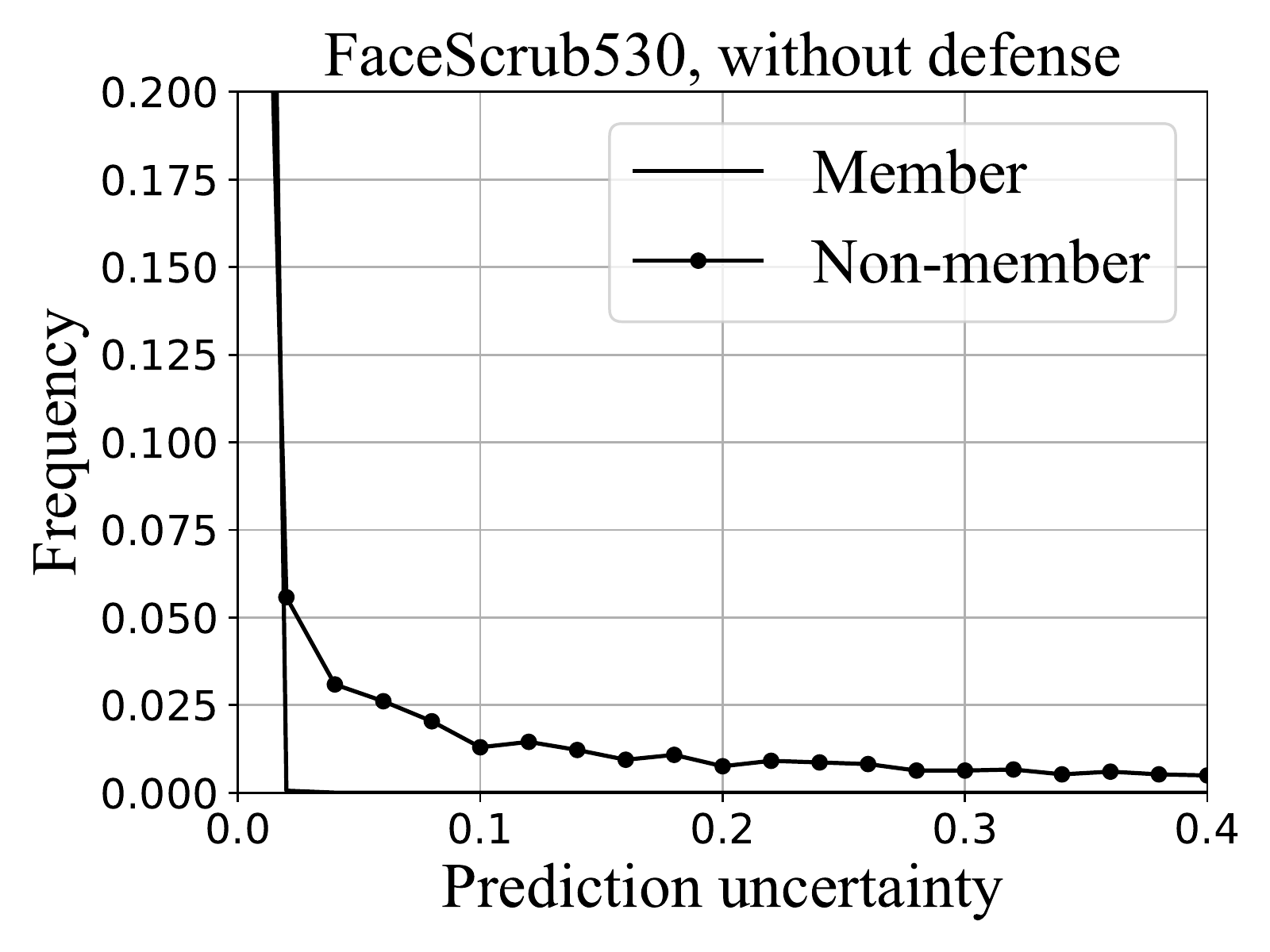}
			\includegraphics[width=0.5\linewidth]{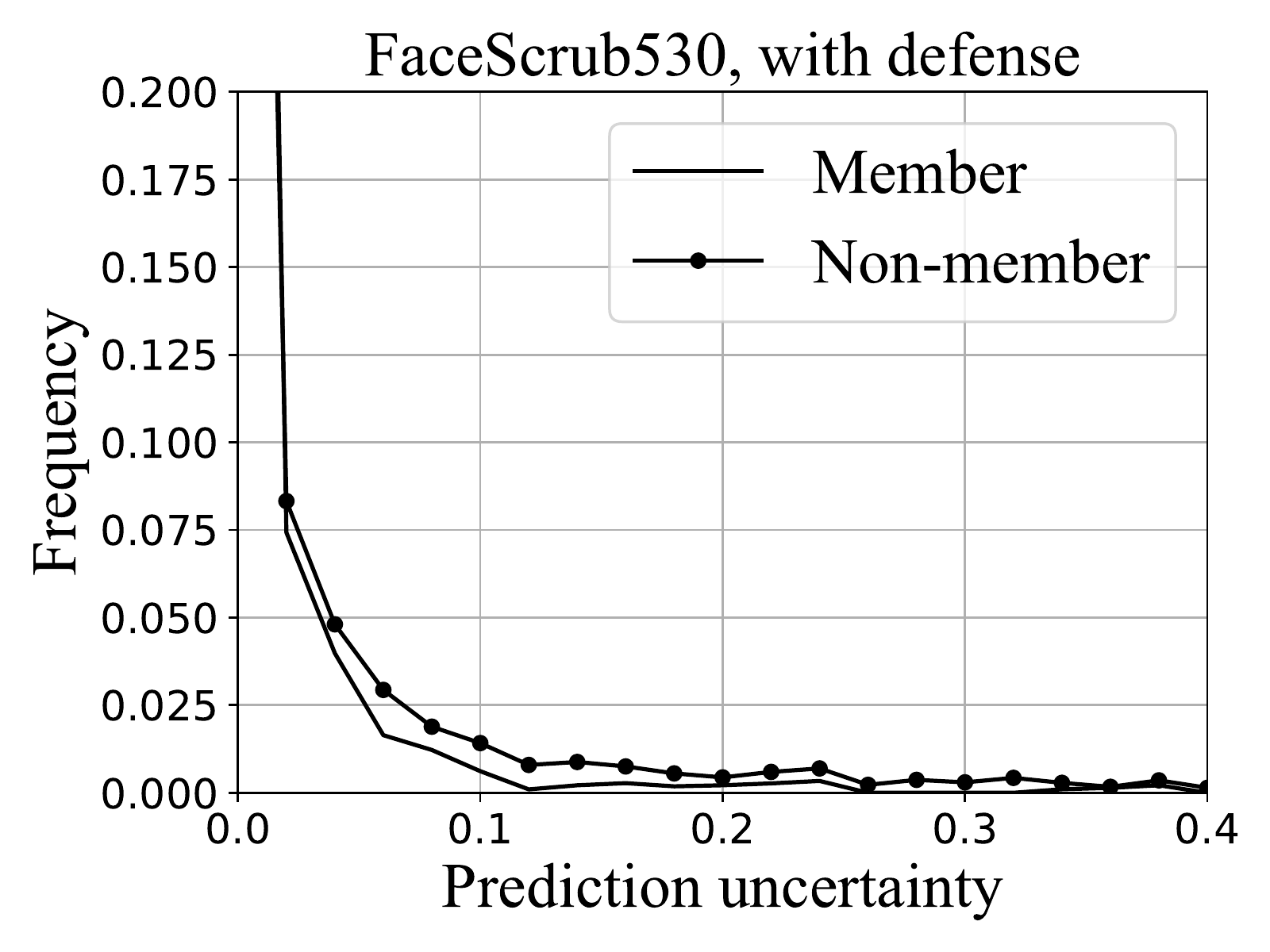}
		}
	\end{minipage}
	\caption{Distribution of the target classifier's prediction uncertainty on members and non-members of its training set. The uncertainty is measured as the normalized entropy of the confidence score vector.}
	\label{fig:uncertainty}
\end{figure}

\begin{table}[t]
	\centering
	\setlength{\tabcolsep}{4pt}
	\caption{Gap of the classifier's confidence in predicting the correct class and the prediction uncertainty between members and non-members.}
	\label{tb:measurement}
	\begin{tabular}{l|l|c|c|c|c|c|c}
		\hline
		\multicolumn{1}{c|}{Measurement} & \multicolumn{1}{c|}{Defense} & \multicolumn{2}{c|}{CIFAR10} & \multicolumn{2}{c|}{Purchase100} & \multicolumn{2}{c}{FaceScrub530} \\
				\hhline{~|~|-|-|-|-|-|-}
		&  & Max & Average & Max & Average & Max & Average \\
		\hline
		\multirow{2}{*}{Confidence} & None & 0.103 & 0.004 & 0.412 & 0.016 & 0.416 & 0.017 \\
		& Purifier & 0.057 & 0.002 & 0.164 & 0.007 & 0.264 & 0.010 \\
		\hline
		\multirow{2}{*}{Uncertainty} & None & 0.114 & 0.005 & 0.201 & 0.015 & 0.418 & 0.017 \\
		& Purifier & 0.019 & 0.002 & 0.058 & 0.003 & 0.129 & 0.005 \\
		\hline
	\end{tabular}
\end{table}

We further investigate how the confidence scores are modified in the purification process. 
We examine the indistinguishability of the confidence scores on members and non-members by plotting the histogram of the target classifier's confidence in predicting the correct class in Figure~\ref{fig:prediction_acc} and also plotting the histogram of the prediction uncertainty in Figure~\ref{fig:uncertainty}.
The prediction uncertainty is measured as the normalized entropy $\frac{-1}{\log (k)} \sum_i \hat{\vec{y}_i}\log (\hat{\vec{y}_i})$ of the confidence vector $\vec{y} = F(\vec{x})$, where $k$ is the number of classes.
In Figure~\ref{fig:prediction_acc} and \ref{fig:uncertainty}, the larger the gap between the two curves is, the more information about the training dataset the model leaks. 
We report the maximum gap and the average gap between the curves (i.e., without versus with defense) in Table~\ref{tb:measurement}.
The results show that our approach can significantly reduce both the maximum and average gaps between the target classifier's confidence in predicting the correct class as well as the prediction uncertainty on its members versus non-members.
This is helpful to reduce what the attacker can exploit to perform the membership inference. 
Besides, they also visibly show how the indistinguishability of the confidence scores on members and non-members can improve by the purification framework.

Figure~\ref{fig:soft_label} visibly presents the confidence score vectors of the Purchase100 classifier on the training and test data from class 29 without and with our defense.
Without defense, the target classifier produces a very high probability for class 29 on the training data. 
On the test data, besides class 29, the probabilities for class 54 and 70 are also similarly high. This is where the target classifier makes most mistakes on the test data. 
Besides, the target classifier spreads the prediction probability across many classes which makes the prediction sensitive to the change of the input data.
When our defense is applied, the prediction probabilities on training data and test data exhibit more similar patterns. For example, besides class 29, they are both high on class 15, 24, 35 and 38. 
The probability of the test data is still high on class 54 and 70 which is in line with the experimental result that our defense preserves the utility.
On the other hand, by limiting the dispersion of the confidence scores, our approach can improve their insensitivity to the change of the input data, which can mitigate the model inversion attack.

\noindent \underline{\textbf{Effect of Adversarial Model}}

\begin{table}[t]
	\centering
	\setlength{\tabcolsep}{6pt}
	\caption{Effect of the adversarial model on the defense performance against the model inversion attack. We use three different adversarial models: $H_1$, $H_2$ and $H_3$. The results are reported on the FaceScrub530 dataset.}
	\label{tb:inversion_only}
	\begin{tabular}{c|c|c|c}
		\hline
		Adversarial model & Classification & Confidence distortion & Inversion error \\
		\hline
		None & 77.68\% & 0 & 0.011448  \\
		\hline
		$H_1$ & 77.48\% & 3.34\% & 0.044825 \\
		\hline
		$H_2$ & 77.52\% & 3.78\% & 0.044876 \\
		\hline
		$H_3$ & 77.48\% & 4.84\% & 0.046276 \\
		\hline
	\end{tabular}
\end{table}

Table~\ref{tb:inversion_only} shows the performance of the purifier in defending the model inversion attack when we use 3 different adversarial models $H_1$, $H_2$ and $H_3$ to train the purifier. 
Specifically, $H_1$ is the original adversarial model. Compared to $H_1$, $H_2$ removes a transposed convolutional layer and its corresponding batch normalization and Tanh activation functions. $H_3$ adds an additional transposed convolutional layer. 
The attacker's inversion model remains unchanged in the experiment. 
We can see that the purifier has similar effectiveness though trained with different adversarial models. The inversion error is significantly increased by a factor of around 4 while confidence score distortion is within 5\% when the purifier is applied.

\noindent \underline{\textbf{Effect of Purifier's Training Data}}

\begin{table}[t]
	\centering
	\setlength{\tabcolsep}{2pt}
	\caption{Effect of the purifier's in-distribution training data on the defense performance. The numbers are reported on Purchase100 dataset.}
	\label{tb:effectreference}
	\begin{tabular}{c|c|c|c|c|c|c}
		\hline
		Training set & Classification & Conf. dist.  & Inver. error & NSH & Mlleaks & Mlleaks-a  \\
		\hline
		$D_2$ (5,000) & 83.15\% & 6.37\% & 0.152346 & 59.60\% & 52.59\% & 54.42\%   \\
		\hline
		$D_2$ (10,000) & 84.47\% & 4.61\% & 0.152175  & 58.50\% & 51.88\% & 54.09\% \\
		\hline
		$D_2$ (20,000) & 84.17\% & 4.38\% & 0.152097 & 58.13\% & 51.31\% & 52.82\%  \\
		\hline
		$D_2$ (40,000) & 84.43\% & 4.18\% & 0.152092 & 58.08\% & 50.90\% & 52.80\%  \\
		\hline
		$D_2$ (60,000) & 84.41\% & 4.00\% & 0.151909 & 58.27\% & 51.21\% & 54.20\%  \\
		\hline
		$D_1$ (20,000) & 81.18\% & 7.91\% & 0.152600 & 59.15\% & 52.77\% & 56.37\%  \\
		\hline
	\end{tabular}
\end{table}

\begin{table}[t]
	\centering
	\setlength{\tabcolsep}{1pt}
	\caption{Effect of the purifier's out-of-distribution training data on the defense performance.}
	\label{tb:effectoutof}
	\begin{tabular}{l|l|l|l|l|l|l|l}
		\hline
		Classifier & Purifier & Classi. & Conf. dist. & Inver. error & NSH & Mlleaks & Mlleaks-a \\
		\hline
		FaceScrub530 & CIFAR10 & 50.30\% & 50.27\% & 0.0430 & 61.50\% & 55.49\% & 59.29\% \\
		\hline
		Purchase100 & Random & 7.68\% & 144.98\% & 0.1619 & 62.99\% & 52.30\% & 53.31\% \\
		\hline
	\end{tabular}
\end{table}

We investigate the effect of the purifier's training data by using different in-distribution data and out-of-distribution data to train the purifier (jointly trained with the adversarial model and the discriminator).
Specifically, for in-distribution data, we vary the size of $D_2$ and also replace $D_2$ with $D_1$ to train the purifier. For out-of-distribution data, we use CIFAR10 data to train the purifier for the FaceScrub530 classifier, and use randomly generated data to train the purifier for the Purchase100 classifier.

We present the effect of the in-distribution training data in Table~\ref{tb:effectreference}.
The results show that the membership inference accuracy (NSH, Mlleaks and Mlleaks-a) decreases as the size of $D_2$ increases from 5,000 to 40,000, but it drops when the size reaches 60,000.
When the size of $D_2$ is relatively small (i.e., 5,000 to 40,000), the purifier learns the most general and important patterns of the confidence score vectors, and such learning is facilitated with more data. Hence, the purified confidence score vectors become less scattered and thus improves their indistinguishability. 
However, when the size of $D_2$ becomes too large (i.e., 60,000), the purifier starts to learn the detailed information of the confidence score vectors. As a result, the purified confidence score vector no longer concentrates on general patterns but becomes an accurate reconstruction, which recovers the distinguishability.
This also explains the observation that the confidence score distortion decreases as the size of $D_2$ increases.
The inversion error has a trend of decrease when the size of $D_2$ increases, but the change is within a very small portion (i.e., 0.2\%). We speculate that this might be due to the same reason that the purified confidence score vector becomes closer to the original one, which is vulnerable to the model inversion attack.

When we use the classifier's training data $D_1$ to train the purifier, it can mitigate the two attacks to some extent but the defense performance is not as good as when it is trained on $D_2$. 
This is mainly because the classification overfitting increases when the purifier is further trained on $D_1$.  As a consequence, the classification accuracy on test set drops by 3\%, and the confidence score distortion increases. The indistinguishability of confidence scores does not improve too much compared with the purifier trained on $D_2$. Therefore, the defense performance against the membership inference is impacted.
Nonetheless, the defense performance against the model inversion attack slightly improves. This might be due to that the adversarial model becomes stronger when trained on overfitted confidence scores, which helps the purifier to better counter the model inversion attack via adversarial learning.

Table~\ref{tb:effectoutof} shows the effect of the out-of-distribution training data. The purifier can still mitigate the model inversion and membership inference attacks, but at the cost of sacrificing the utility of the target classifier. This is not surprising because the purifier cannot extract useful patterns from the confidence scores on out-of-distribution data, which makes the purified confidence information meaningless.

\subsection{Comparison with Other Methods }

\begin{table}[t]
	\setlength{\tabcolsep}{0.5pt}
	\centering
	\caption{Efficiency of different defense methods.}
	\label{tb:efficiency}
	\begin{tabular}{l|c|c|c|c|c|c|c|c}
		\hline
		\multicolumn{1}{c|}{Defense} & \multicolumn{2}{c|}{CIFAR10} & \multicolumn{2}{c|}{Purchase100} & \multicolumn{2}{c|}{FaceScrub530} & \multicolumn{2}{c}{Average} \\
		\hhline{~|-|-|-|-|-|-|-|-}
		& Train & Test & Train & Test & Train & Test & Train & Test\\
		\hline
		None & 100\% & 100\% & 100\% & 100\% & 100\% & 100\% & 100\% & 100\%\\
		\hline
		Purifier & 107\% & 128\% & 703\% & 214\% & 347\% & 150\% & 386\% & 164\%\\
		\hline
		Min-Max & 513\% & 103\% & 2,798\% & 100\% & 251\% & 102\% & 1,187\%& 102\% \\
		\hline
		MemGuard & 113\% & 635,000\% & 287\% & 1,120,000\% & 133\% & 526,000\% & 178\% & 760,333\%\\
		\hline
		Model-Stack. & 155\% & 192\% & 206\% & 297\% & 253\% & 278\% & 205\%& 256\%\\
		\hline
	\end{tabular}
\end{table}

\begin{figure*}[t]
	\centering
	\begin{minipage}[b]{1\linewidth}
		\centering
		\subfigure{
			\includegraphics[width=\linewidth]{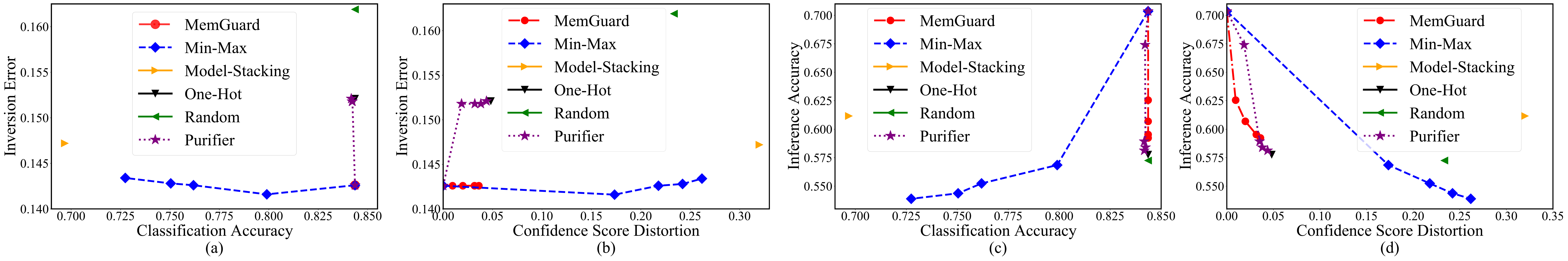}
		}
	\end{minipage}
	\caption{Comparison of different defenses on the Purchase100 dataset. (a) Inversion error vs. classification accuracy. (b) Inversion error vs. classification score distortion. (c) Inference accuracy vs. classification accuracy. (d) Inference accuracy vs. confidence score distortion.}
	\label{fig:compare_purchase}
\end{figure*}

We compare our approach with existing defense approaches listed in Section~\ref{sec:existing}. MemGuard, Min-Max and Purifier have hyper-parameters to control the security-utility tradeoff. For instance, the confidence score distortion budget $\epsilon$ in MemGuard, the hyper-parameter $\lambda$ that balances between the loss function and the adversarial regularizer in Min-Max, and the $\lambda$, $\alpha$ and $\beta$ in Purifier. Model-Stacking, One-Hot Encoding and Random Noise do not have such a hyper-parameter to easily control the tradeoff~\cite{jia_memguard_2019}. The hyper-parameters that we use for each defense are described in Appendix~\ref{app:compare}.

For each defense and a given hyper-parameter (if applicable), we collect 4 pairs of data: (inversion error, classification accuracy), (inversion error, confidence score distortion), (inference accuracy, classification accuracy) and (inference accuracy, confidence score distortion). The first two pairs measure the performance against the model inversion attack, and the last two pairs measure the performance against the membership inference attack (we consider NSH attack). By exploring different hyper-parameters, we can obtain 4 sets of such pairs. We plot each set of pairs in a graph and show them in Figure~\ref{fig:compare_purchase}.
Our results show that Purifier has the best security-utility tradeoff for the model inversion attack. 
Given the same confidence score distortion, it achieves the largest inversion error. 
MemGuard and Min-Max have very limited defense effect.
One-Hot Encoding, Random Noise and Model-Stacking can mitigate the model inversion attack to some extent, but they incur intolerable confidence score distortion.
Figure~\ref{fig:facescrub} visibly compares each defense against the model inversion attack.
For the membership inference attack, Purifier achieves a better security-utility tradeoff than Min-Max, Model-Stacking, One-Hot Encoding and Random Noise.
MemGuard slightly outperforms Purifier when the confidence score distortion is very small. However, Purifier can achieve a smaller inference accuracy than MemGuard by sacrificing more distortion to the confidence scores.
Purifier incurs negligible classification accuracy loss. 
We can obtain similar results on other datasets as shown in Appendix~\ref{app:compare} (Figure~\ref{fig:compare_cifar} and \ref{fig:compare_facescrub}).

We visualize the confidence score vectors produced by each defense in Figure~\ref{fig:tsne}.
It is clear that Purifier transforms the confidence score vectors to have very clear patterns. Their dispersion is also very small.
As a result, Purifier can concurrently defend both attacks. 
MemGuard, Min-Max and Model-Stacking do not have such impact, so they only work for the membership inference attack. 
Although One-Hot Encoding and Random Noise can mitigate both attacks without transforming the confidence scores in the Purifier's way, they lose all the confidence information in the prediction and thus cannot work as practical solutions when the confidence is required.

We report the efficiency of each defense in Table~\ref{tb:efficiency}. We perform our experiments on a PC equipped with four Titan XP GPUs with 12GBytes of graphic memory,128 GBytes of memory and an Intel Xeon E5-2678 CPU. 
Due to the training of three additional models (purifier, adversarial model and discriminator) in Purifier, its training time is 2.86 times more than the target classifier, which is considered acceptable as a one-time effort.
The testing time of Purifier is 1.64 times as much as the target classifier, which is considered comparable with Min-Max and Model-Stacking. 
However, the testing time of MemGuard is 7,000+ times more than the original classifier, which introduces significant efficiency overhead.

\section{Discussion}

Involving an in-distribution reference dataset in the defense mechanism is common in the literature. For instance, MemGuard uses a reference set to train the defense classifier~\cite{jia_memguard_2019}. Min-Max uses it to train the inference model~\cite{nasr_machine_2018}. Similarly, our approach uses it to train the purifier. 
Unfortunately, little has been discussed about whether such reference dataset brings vulnerability for data inference attacks.
Assuming the reference data are considered as members, we present the inversion error and the inference accuracy (we consider NSH attack) on the reference set $D_2$ and the test set $D_3$ for each defense in Table~\ref{tb:attack_d2}. 
Results show that the inference accuracy does not increase on the reference set compared with the original training data of the target classifier.
Purifier can still preserve the defense effect against the model inversion attack.
However, there might be opposing views on whether such reference dataset should be considered as members.

\begin{table}[t]
	\centering
	\setlength{\tabcolsep}{10pt}
	\caption{Results of model inversion attack and membership inference attack on the reference set for different defenses. The experiments are performed on FaceScrub530 dataset.}
	\label{tb:attack_d2}
	\begin{tabular}{c|c|c}
		\hline
		Defense & Inversion error & Inference accuracy\\
		\hline
		Purifier \rom{1} \& \rom{2} & 0.0450 & 53.63\%\\
		\hline
		Min-Max & 0.0222 & 52.07\% \\
		\hline
		MemGuard & 0.0116 & 52.20\% \\
		\hline
	\end{tabular}
\end{table}

\section{Related Work}

\textbf{Inference Attacks.}
The inference attacks against machine learning can be divided into model inference and data inference attacks. In model inference attacks, an attacker could infer the parameters~\cite{tramer_stealing_2016}, hyper-parameters~\cite{wang_stealing_2018}, architecture~\cite{oh_towards_2018} and functionality~\cite{orekondy_knockoff_2019} of a target model.
In data inference attacks, the attacker aims at inferring information about the data that the target model operates on. These attacks include membership inference attack~\cite{shokri_membership_2017}, model inversion attack (input inference)~\cite{yang_neural_2019, fredrikson_model_2015}, attribute inference~\cite{fredrikson_privacy_2014, wu_methodology_2016}, statistics inference~\cite{ateniese_hacking_2015} and side-channel attack~\cite{wei_i_2018}. 

In this paper, we concentrate on data inference attacks, notably membership inference attack and model inversion attack. Existing defenses mainly focused on membership inference attacks~\cite{shokri_membership_2017, nasr_machine_2018, jia_memguard_2019, li_membership_2020}. Little has been studied about the model inversion attack.
Xiao et al.~\cite{xiao_adversarial_2019} studied the adversarial reconstruction problem where they aimed to prevent the latent representations from being decoded into the original input data. 
To this end, they regularized the encoder with an adversarial loss from a decoder. 
They studied the face attribute prediction model which outputs 40 binary facial attributes. 
Our paper, on the contrary, studies black-box classification models whose output is constrained by a probability distribution wherein the values sum up to 1. Moreover, they did not consider the adversarial scenario
where the attacker has no access to the same data distribution as the original training data.
Jia and Gong~\cite{jia_attriguard_2018} proposed the adversarial formulation for privacy protection. They aimed at protecting the privacy of users' sensitive attributes from being inferred from their public data. Our work, on the other hand, investigates inference attacks that leverage the prediction results of machine learning models to infer useful information about the input data.

\textbf{General Membership Inference Attack.}
Membership inference attack is performed to determine whether a given data sample is part of a target dataset. It is not limited to machine learning models. Homer et al.~\cite{homer_resolving_2008} proposed one of the first membership inference attacks in the biomedical setting on genomic data. Some studies also performed membership inference attacks on other biomedical data such as MicroRNA~\cite{backes_membership_2016} and DNA methylation~\cite{hagestedt_mbeacon_2019}. Pyrgelis et al.~\cite{pyrgelis_knock_2018, pyrgelis_under_2019} further showed that it is possible to perform membership inference attack on location datasets as well.
Shokri et al.~\cite{shokri_membership_2017} performed membership inference attack in the machine learning setting which is the same with our work.

\textbf{Secure \& Privacy-Preserving Machine Learning.}
The untrusted access of machine learning models in the training or testing phase is a direct threat to the security and privacy of machine learning. A number of studies made use of trusted hardware and cryptographic computing to provide secure and privacy-preserving training and use of machine learning models. 
These techniques include homomorphic encryption, garbled circuits and secure multi-party computation on private data~\cite{liu_oblivious_2017, bonawitz_practical_2017, phong_privacy-preserving_2018, dowlin_cryptonets_2016, mohassel_secureml_2017, dwork_privacy-preserving_2018} and secure computing using trusted hardware~\cite{ohrimenko_oblivious_2016, juvekar_gazelle_2018}.
Although these methods protect sensitive data from direct observation by the attacker, they do not prevent information leakage via the model computation itself which could be exploited by various inference attacks.

\section{Conclusion}

We propose a purification framework to defend the model inversion attack and the membership inference attack.
The framework purifies confidence score vectors by reducing their dispersion.
It can be further specialized in defending a particular attack by adversarial learning.
Our extensive experiments show that the purification framework is effective in mitigating the two attacks while imposing negligible utility loss. We also show that training the purifier for defending one attack can also mitigate the other one, which empirically show the connection of the two attacks.


\appendix

\subsection{Datasets}
\label{app:dataset}

\noindent \textbf{CIFAR10~\cite{shokri_membership_2017, salem_ml-leaks_2018, li_membership_2020}.}
It is a machine learning benchmark dataset for evaluating image recognition algorithms. 
It consists of 60,000 color images, each of size 32 x 32. The dataset has 10 classes, where each class represents an object (e.g., airplane, car, etc.)

\noindent \textbf{Purchase100~\cite{shokri_membership_2017, nasr_machine_2018, salem_ml-leaks_2018, li_membership_2020}.}
This dataset is based on Kaggle's ``acquired valued shopper'' challenge\footnote{https://www.kaggle.com/c/acquire-valued-shoppers-challenge/data}. 
We used the preprocessed and simplified version of this dataset~\cite{shokri_membership_2017}.
It is composed of 197,324 data records and each data record has 600 binary features. The dataset is clustered into 100 classes.

\noindent \textbf{FaceScrub530~\cite{yang_neural_2019}.}
This dataset consists of URLs for 100,000 images of 530 individuals.
We obtained the preprocessed and simplified version of this dataset from \cite{yang_neural_2019} which has 48,579 facial images and each image is resized to 64 $\times$ 64.

\subsection{Purification Framework Model Architecture}
\label{app:purification}
\noindent \textbf{Purifier.} We use an autoencoder structure to implement the purifier. It has the layer size [10, 7, 4, 7, 10] for CIFAR10, [100, 50, 20, 10, 20, 50, 100] for Purchase100 and [530, 200, 530] for FaceScrub530.
We use the ReLU activation function and batch normalization in every hidden layer of all the purifier models.
We train the Purchase100 purifier for 200 epochs and others for 50 epochs. 
We use the Adam optimizer with learning rate 0.01 for CIFAR10, 0.0001 for Purchase100 and 0.0005 for Facescrub530.

\noindent \textbf{Adversarial Model.} We use different adversarial models for different datasets. For FaceScrub530, we use the same model as \cite{yang_neural_2019}, which consists of 5 transposed CNN blocks. 
For CIFAR10, we reduce 1 transposed CNN blocks from the FaceScrub530 adversarial model and the model consists of 4 transposed CNN blocks. 
For Purchase100, we use a multi-layer perceptron of size [100, 512, 1024, 600].
Each hidden layer has a ReLU activation function and the output layer has a Sigmoid activation function.
We use the Adam optimizer with learning rate 0.0002 for all the adversarial models.

\noindent \textbf{Discriminator.} We use a similar structure as the attack model in \cite{nasr_machine_2018} to build the discriminator. Specifically, it consists of three neural networks. The first neural network operates on the confidence score vector and has the size of [$d$, 1024, 512, 64] where $d$ is the input dimension.
The second neural network takes the one-hot encoded label as input and has the size of [$d$, 512, 64].
The third network takes the concatenation of the output of the first two networks as input and produces a single value indicating whether the input confidence score vector and label are real or fake.
We use the Adam optimizer with learning rate 0.0002 to train the discriminator model.

\subsection{Comparison  on Existing Defenses}
\label{app:compare}

\begin{figure*}[t]
	\centering
	\begin{minipage}[b]{1\linewidth}
		\centering
		\subfigure{
			\includegraphics[width=\linewidth]{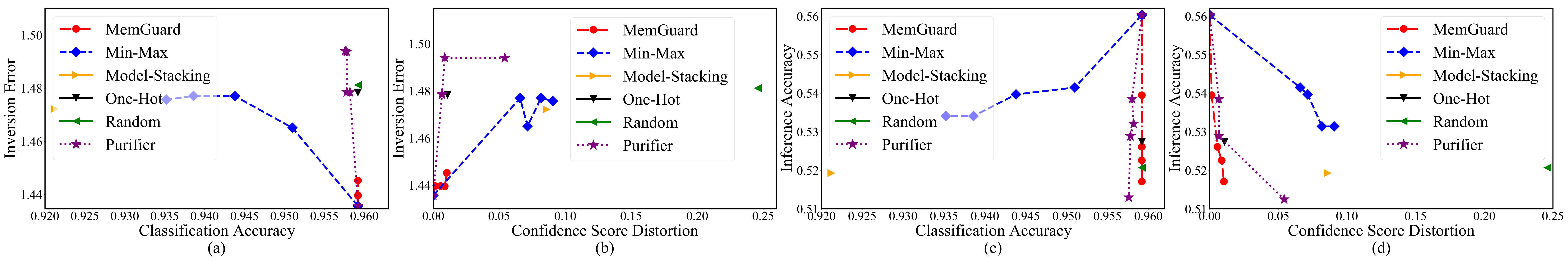}
		}
	\end{minipage}
	\caption{Comparison of different defenses on the CIFAR10 dataset. (a) Inversion error vs. classification accuracy. (b) Inversion error vs. classification score distortion. (c) Inference accuracy vs. classification accuracy. (d) Inference accuracy vs. confidence score distortion.}
	\label{fig:compare_cifar}
\end{figure*}

\begin{figure*}[t]
	\centering
	\begin{minipage}[b]{1\linewidth}
		\centering
		\subfigure{
			\includegraphics[width=\linewidth]{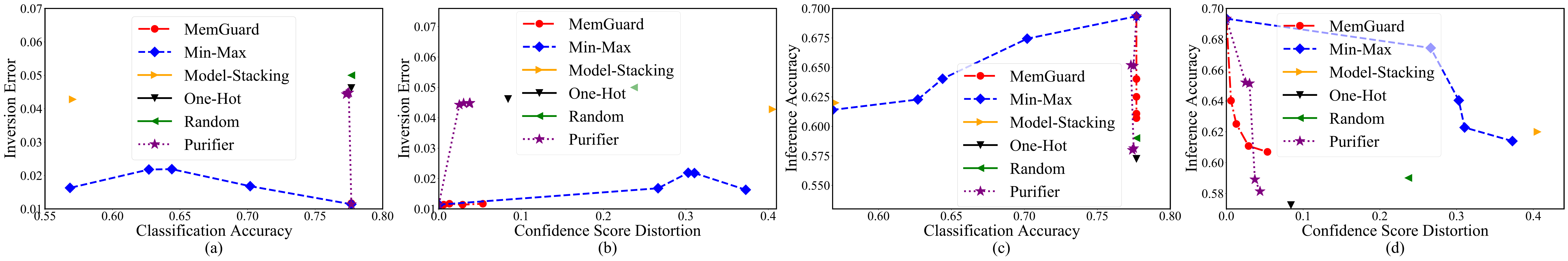}
		}
	\end{minipage}
	\caption{Comparison of different defenses on the FaceScrub530 dataset. (a) Inversion error vs. classification accuracy. (b) Inversion error vs. classification score distortion. (c) Inference accuracy vs. classification accuracy. (d) Inference accuracy vs. confidence score distortion.}
	\label{fig:compare_facescrub}
\end{figure*}

We tried the hyper-parameter $\lambda$ of Min-Max Game in the range [2,10] with a step size of 2.0 for Purchase100 and FaceScrub530 and [5,25] with a step size of 5.0 for CIFAR10. 
We tried the parameter $\epsilon$ that controls the level of noise in [0,0.1,0.2,0.4,0.7,1.0] in the implementation of MemGuard. 
As for Purifier,we choose $\lambda$ from [0.1,1,5] while $\alpha$ and $\beta$ are from [0,0.1,1,5,10] for each dataset.
We show the comparison of different defenses on the CIFAR10 dataset in Figure~\ref{fig:compare_cifar}, and on the FaceScrub dataset in Figure~\ref{fig:compare_facescrub}.

\end{document}